\documentclass[aps,pre,superscriptaddress, longbibliography,twocolumn,floatfix,noeprint]{revtex4-2}

\usepackage{amsmath}
\usepackage{amssymb}
\usepackage{sansmath}
\usepackage{amsfonts}
\usepackage{graphicx}
\usepackage{dcolumn}
\usepackage{bm}
\usepackage{latexsym}
\usepackage{slashed}
\usepackage{xcolor}
\usepackage[colorlinks=true,linkcolor=blue,citecolor=blue]{hyperref}
\usepackage{array}

\begin{document}

\title{Quasistatic modeling of ultrarelativistic beam-plasma instabilities}

\author{P.~San Miguel Claveria}
\thanks{These authors have contributed equally to this work.}
\affiliation{GAP/Instituto de Plasmas e Fusão Nuclear, Instituto Superior Técnico, Universidade de Lisboa, Lisbon, 1049-001, Portugal}
\affiliation{Laboratoire d’Optique Appliquée (LOA), CNRS, École polytechnique, ENSTA, Institut Polytechnique de Paris, Palaiseau, France}

\author{L.~Gremillet}
\thanks{These authors have contributed equally to this work.}
\affiliation{CEA, DAM, DIF, F-91297 Arpajon, France}
\affiliation{Universit\'{e} Paris-Saclay, CEA, LMCE, F-91680 Bruy\`{e}res-le-Ch\^{a}tel, France}

\author{X.~Davoine}
\affiliation{CEA, DAM, DIF, F-91297 Arpajon, France}
\affiliation{Universit\'{e} Paris-Saclay, CEA, LMCE, F-91680 Bruy\`{e}res-le-Ch\^{a}tel, France}

\author{Q.~Labro}
\affiliation{GAP/Instituto de Plasmas e Fusão Nuclear, Instituto Superior Técnico, Universidade de Lisboa, Lisbon, 1049-001, Portugal}

\author{A.~Matheron}
\altaffiliation[Present address: ]{Helmholtz-Institut Jena, Fr\"obelstieg 3, 07743 Jena, Germany}
\affiliation{Laboratoire d’Optique Appliquée (LOA), CNRS, École polytechnique, ENSTA, Institut Polytechnique de Paris, Palaiseau, France}

\author{M.~Tamburini}
\affiliation{Max-Planck-Institut f\"ur Kernphysik, Saupfercheckweg 1, D-69117 Heidelberg, Germany}

\author{F.~Fiuza}
\affiliation{GAP/Instituto de Plasmas e Fusão Nuclear, Instituto Superior Técnico, Universidade de Lisboa, Lisbon, 1049-001, Portugal}

\author{S.~Corde}
\email[Corresponding authors:\\ ]{pablo.san.miguel.claveria@tecnico.ulisboa.pt\\ laurent.gremillet@cea.fr\\ frederico.fiuza@tecnico.ulisboa.pt\\ sebastien.corde@polytechnique.edu}
\affiliation{Laboratoire d’Optique Appliquée (LOA), CNRS, École polytechnique, ENSTA, Institut Polytechnique de Paris, Palaiseau, France}

\date{\today}

\begin{abstract}
Relativistic particle beams propagating through dense ambient plasmas are susceptible to streaming instabilities that can govern the system dynamics in various astrophysical and laboratory settings. For an unmagnetized, collisionless plasma pervaded by a dilute, cold relativistic beam, the dominant instabilities are the quasielectrostatic, oblique two-stream (OTSI) and the essentially magnetic, current filamentation instability (CFI). While their linear and nonlinear properties have been researched for decades, most treatments assume unbounded, uniform systems and thus predict purely temporal instability growth---i.e., at a rate independent of position along the beam. This assumption, however, is questionable for realistic configurations where a bounded beam continuously encounters ``fresh'' plasma. This feature causes instabilities to grow in a spatiotemporal manner and, for short beams, at a much slower rate than predicted by temporal theory. Whereas spatiotemporal perturbative treatments of streaming instabilities were derived as early as the 1960s, only recently have the spatiotemporal regimes of OTSI and CFI been addressed theoretically. Yet these models are restricted to a specific instability class, and hence cannot describe the competition between spatiotemporal OTSI and CFI. In this work, we present a unified, fully electromagnetic model of all unstable modes arising throughout the beam. This model leverages the quasistatic approximation to decouple the slow evolution of high-inertia beam particles from the fast dynamics of background plasma electrons. Unlike previous studies, we do not adopt the slowly varying envelope approximation (SVEA), which enables us to track dominant modes across all spatial scales, including the vicinity of the beam front.
Under conditions where standard theory predicts dominance of temporal OTSI, we find that a previously unreported spatiotemporal CFI actually prevails near the front---precisely where the SVEA fails---and is only superseded by OTSI further back in the beam. Furthermore, we demonstrate that our model also captures the growth of the self-modulation and hosing instabilities excited by long, narrow beams, highlighting the fundamental connection between these longitudinal modes and their transverse counterparts. Comparisons with particle-in-cell simulations confirm the validity of the quasistatic approach for modeling streaming plasma instabilities triggered by relativistic dilute beams.
\end{abstract}

\maketitle                                                                                              
\section{Introduction}
The interaction of relativistic particle beams with dense background plasmas is a ubiquitous phenomenon in high-energy astrophysical and laboratory environments. In weakly collisional regimes, this interaction is mediated by streaming instabilities that amplify electromagnetic field modulations---coupled with particle density and current perturbations---to the point of governing the system's energy and momentum transfers \cite{Sudan_Handbook_1984, Bret_POP_2010}. Under initially unmagnetized conditions, and for a transverse beam extent much larger than the background plasma's skin depth ($k_p^{-1}\equiv c/\omega_p$, where $\omega_p$ is the background plasma frequency and $c$ the velocity of light), these instabilities are categorized by the orientation of their wavenumber relative to the mean beam velocity: longitudinal two-stream (TSI), oblique two-stream (OTSI), and perpendicular current filamentation (CFI) modes. These modes further differ in their growth rates, phase velocities, and electromagnetic nature, ranging from electrostatic (TSI) and quasielectrostatic (OTSI) to primarily magnetic (CFI) \cite{Bret_PRL_2008, Bret_POP_2010, Lemoine_MNRAS_2010}. Consequently, they act upon the beam and plasma populations in distinct ways during their evolution.

In astrophysics, these collective processes underpin the microphysics of relativistic jets in gamma-ray bursts, pulsar wind nebulae, and active galactic nuclei. They are thought to generate Fermi-accelerating collisionless shock waves \cite{Medvedev_APJ_1999, Spitkovsky_APJ_2008a, Martins_APJ_2009, Lemoine_PRL_2019, Groselj_APJL_2024} and drive the magnetized turbulence that enables high-energy synchrotron radiation \cite{Sironi_APJL_2009, Medvedev_APJ_2011}. A subject of active debate is whether they can modify the propagation of blazar-induced TeV electron-positron pair beams through the intergalactic medium sufficiently to compete with (inverse Compton) radiative cooling in the cosmic microwave background \cite{Broderick_APJ_2012, Schlickeiser_APJ_2012, Sironi_APJ_2014, Rafighi_AA_2017, Vafin_APJ_2018}.

In the laboratory, these instabilities can disrupt the propagation of the particle beams driving plasma-wakefield accelerators \cite{Katsouleas_PRA_1986}. They are also central to the generation and transport of suprathermal electrons in relativistic laser-matter interactions involving both overcritical \cite{Sentoku_PRL_2003, Jung_PRL_2005, Adam_PRL_2006, Fiuza_PRL_2012, Robinson_NF_2014, Heron_POP_2015, Huang_POP_2017} and undercritical \cite{Huntington_PRL_2011} plasmas. Their potential evolution into collisionless shocks of astrophysical relevance has been explored numerically
\cite{Fiuza_PRL_2012, Ruyer_POP_2015, Grassi_PRE_2017}, while recent proposals suggest harnessing them to mediate intense gamma-ray emissions \cite{Benedetti_NP_2018, Gong_PRL_2023} in accelerator-based experiments.

Over the past decade or so, dedicated experiments have investigated these instabilities under better-controlled conditions at accelerator facilities using moderately relativistic electron beams \cite{Allen_PRL_2012, Gross_PRL_2018}, ultrarelativistic proton beams \cite{Verra_PRE_2024}, and electron-positron beams \cite{Arrowsmith_NC_2024, Arrowsmith_PNAS_2025, Halliday_Arxiv_2026}. Similar efforts using ultraintense lasers have also been undertaken \cite{Gode_PRL_2017, Raj_PRR_2020, Ordyna_NC_2024, Schoenwaelder_NC_2026}, albeit typically with less precise parameter control.

Theoretically, while these phenomena have been extensively studied since the 1960s \cite{Bludman_POF_1960, Fainberg_JETP_1970, Breizman_JETP_1971, Lee_PRL_1973, Thode_POF_1975a, Molvig_PRL_1975, Califano_PRE_1997, Silva_POP_2002, Bret_POP_2010, Lemoine_MNRAS_2010}, most existing analytical and numerical models assume unbounded, uniform systems subject only to temporal instabilities---i.e., instabilities growing at a rate independent of spatial position. Yet, this assumption is at odds with realistic configurations, where a bounded beam continuously interacts with unperturbed plasma at its leading edge \cite{Shukla_NJP_2020}. This discrepancy was addressed early on via more involved perturbative treatments accounting for the coupled spatial and temporal evolution of the instabilities in such systems \cite{Briggs_book_1964, Evans_POF_1970, Bers_Handbook_1983, Jones_POF_1983, Bers_PRL_1984, Rostomyan_POP_2000}. These studies, though, did not address the quasielectrostatic OTSI that largely dominates TSI in the relativistic cold-beam regime, nor did they address the primarily magnetic CFI that develops normal to the beam direction.

These limitations were addressed in our previous work~\cite{San_Miguel_Claveria_PRR_2022}, where a spatiotemporal model for OTSI was developed. A major finding was that spatiotemporal effects dominate OTSI for relativistic beams shorter than $\sim c/\Gamma_{\rm OTSI}$ (where $\Gamma_{\rm OTSI}$ is the temporal OTSI growth rate \cite{Bludman_POF_1960, Bret_POP_2010}), leading to a significantly reduced growth rate compared to unbounded systems. This model, however, is restricted to the electrostatic limit and employs the slowly varying envelope approximation (SVEA), which fails near the beam front. Conversely, a spatiotemporal treatment of CFI was proposed in Ref.~\cite{Pathak_NJP_2015} under the SVEA and the assumption of fully inductive modes. That study predicted that, in the relativistic regime, spatiotemporal effects are confined so close to the beam front that standard temporal growth effectively characterizes the overall CFI dynamics.

Here, we move beyond these shortcomings by presenting a unified, fully electromagnetic, perturbative model that remains valid even in the immediate vicinity of the beam front. This model relies on the quasistatic approximation (QSA), which decouples the slow dynamics of the relativistic beam particles from the fast response of the plasma electrons---a simplification widely used in plasma accelerator studies \cite{Sprangle_PRA_1990, Mora_POP_1997}. This approach allows us to derive a governing equation for the linear growth phase of all unstable modes, whether (quasi)electrostatic or electromagnetic, and thus to describe their early spatiotemporal interplay within the beam. Importantly, by avoiding the SVEA, we can describe the development of unstable modes across all spatiotemporal scales.

Under conditions that standard theory predicts are governed by temporal OTSI, our new theoretical framework reveals two key behaviors: (i) spatiotemporal CFI prevails in a narrow region near the beam front (over $\xi \lesssim 0.385\,\Gamma_{\rm CFI}\tau$, where $\Gamma_{\rm CFI}$ is the temporal CFI growth rate and $\tau$ is the interaction time), yet much more extended than previously thought in the relativistic regime \cite{Pathak_NJP_2015}; (ii) spatiotemporal OTSI takes over deeper into the beam ($\xi \gtrsim 0.385\,\Gamma_{\rm CFI}\tau$). These predictions are corroborated by high-resolution particle-in-cell (PIC) simulations. 

Furthermore, we apply our general quasistatic theory of beam-plasma instabilities to the longitudinal modes associated with TSI in a transversely wide and uniform beam, as well as to the self-modulation (SMI~\cite{Verra_PRL_2022, Verra_POP_2023}) and hosing (HI~\cite{Schroeder_PRE_2012}) instabilities excited by beams with narrow transverse widths $\sigma_y \ll k_p^{-1}$. Our unified model recovers the dominant spatiotemporal evolution of each of these modes during their linear stage, highlighting the fundamental relationship between SMI/HI and OTSI in the limit of an infinitely large transverse wavenumber.

The paper is organized as follows. In Sec.~\ref{sec:QSA_model}, we introduce the quasistatic fluid model describing the coupled evolution of the beam and plasma electrons, which yields a general differential equation for the spatiotemporal evolution of the full unstable electromagnetic spectrum. Section~\ref{sec:transverse_modes} applies this model to transverse modes (OTSI and CFI), first deriving analytical asymptotic solutions for longitudinally semi-infinite, flat-top beams, and then numerically extending these solutions to Gaussian longitudinal beam profiles. In Sec.~\ref{sec:Longitudinal_modes}, the formalism is extended to longitudinal modes, focusing first on TSI for wide beams and subsequently on narrow-beam regimes (SMI and HI). High-resolution PIC simulations are presented in both Sec.~\ref{sec:transverse_modes} and Sec.~\ref{sec:Longitudinal_modes} to validate the leading-term analytical solutions. Finally, Sec.~\ref{sec:conclusions} summarizes our main findings and outlines directions for future work.

\section{Quasistatic theory of beam-plasma instabilities}
\label{sec:QSA_model}

In this section, we derive a fully electromagnetic, perturbative theory of spatiotemporal beam-plasma instabilities, employing the quasistatic and cold-fluid approximations. The beam may consist of several particle species, all traveling at the same relativistic velocity $\mathbf{v_b} = v_b \mathbf{\hat{x}}$, with the corresponding Lorentz factor $\gamma_b = 1/\sqrt{1-(v_b/c)^2} \gg 1$ ($c$ is the velocity of light). Each beam species $s$ is characterized by its mass $m_s$, charge $q_s$ and number density $n_{bs}$. The background plasma electrons (mass $m_e$, charge $-e$) are initially at rest, with a number density $n_p \gg n_{bs}$, while the plasma ions are treated as immobile.

Introducing the electron plasma frequency $\omega_p = \sqrt{n_p e^2/m_e \epsilon_0}$, we normalize time as $t \to \omega_p t$, position as $\mathbf{r} \to \mathbf{r}\omega_p/c$, charge as $q_s \to q_s/e$, mass as $m_s \to m_s/m_e$, velocity as $\mathbf{v} \to \mathbf{v}/c$, density as $n \to n/n_p$, the electric field as $\mathbf{E} = e\mathbf{E}/m_e \omega_p c$, the magnetic field as $\mathbf{B} = e\mathbf{B}/m_e \omega_p$, the scalar potential as $\phi \to e\phi/m_e c^2$, and the vector potential as $\mathbf{A} \to e\mathbf{A}/m_e c$. Equilibrium and first-order perturbed quantities are denoted by superscripts $^{(0)}$ and $^{(1)}$, respectively. For simplicity, our analysis is restricted to the $x$--$y$ plane, but it can be readily generalized to a 3D geometry.

In this framework, the linearized momentum and continuity equations for the background plasma electrons read
\begin{align}
    &\partial_t \mathbf{v}_p^{(1)} = - \mathbf{E}^{(1)} \,,
    \label{eq:plasma_eq_motion} \\
    &\partial_t n_p^{(1)} + \mathbf{\nabla} \cdot \mathbf{v_p}^{(1)} = 0 \,,
    \label{eq:plasma_continuity}
\end{align}
assuming that the plasma electron velocity remains nonrelativistic ($v_p \ll 1$).

By taking the time derivative of the continuity equation and combining it with the linearized Gauss's law, 
\begin{equation}
    \mathbf{\nabla} \cdot \mathbf{E}^{(1)} = \sum_s q_s n_{bs}^{(1)} - n_p^{(1)} \,,
\end{equation}
we obtain
\begin{equation}
    \left( \partial^2_t +1 \right) n_p^{(1)} = \rho_b^{(1)} \,,
    \label{eq:plasma_oscillations}
\end{equation}
where $\rho_b = \sum_s q_s n_{bs}$ is the total beam charge density.

We now transform to comoving coordinates $(\xi,\tau) = (t-x,t)$. The QSA hinges on the observation that the plasma and field quantities vary much faster in $\xi$ than in $\tau$ \cite{Chen_IEEE_1987}. Consequently, we neglect $\partial_\tau$ relative to $\partial_\xi$ for these quantities, simplifying the above to
\begin{equation}
    \left( \partial^2_\xi +1 \right) n_p^{(1)} = \rho_b^{(1)} \,.
    \label{eq:plasma_oscillations_QSA}
\end{equation}
This equation describes the linear plasma response to perturbations in the total beam charge density~\cite{Chen_PRL_1985}.

The beam species obey the following linearized relativistic equations of motion: 
\begin{align}
    \label{eq:beam_continuity_1}
    &\left( \partial_t + \partial_x \right) n_{bs}^{(1)}+ n_{bs}^{(0)}\left( \partial_x v_{bs,x}^{(1)} + \partial_y v_{bs,y}^{(1)} \right) 
    = 0 \,, \\
\begin{split}
& \left(\partial_t + \partial_x \right) 
\begin{pmatrix}
   		 \gamma_b^{(0)3} v_{bs,x}^{(1)} \\
  		 \gamma_b^{(0)} v_{bs,y}^{(1)}
   	\end{pmatrix}	
    \\ & =  -\frac{q_s}{m_s} 
	\begin{pmatrix}
   		 \partial_x \phi^{(1)} + \partial_t A_x^{(1)} \\
  		 \partial_y (\phi^{(1)} - A_x^{(1)}) + (\partial_t + \partial_x) A_y^{(1)}
   	\end{pmatrix}
    \,. 
    \label{eq:beam_momentum_1}
\end{split}
\end{align}
An initially uniform beam profile ($\partial_\xi n_{bs}^{(0)} =0$ for $\xi >0$) is assumed  in Eq.~\eqref{eq:beam_continuity_1}. The case of a nonuniform beam profile is addressed in Sec.~\ref{sec:gaussian_beams}.

Transforming to $(\xi,\tau)$ and applying the QSA to the fields, we find
\begin{align}
    &\partial_\tau n_{bs}^{(1)} = n_{bs}^{(0)}\left( \partial_\xi v_{bs,x}^{(1)} - \partial_y v_{bs,y}^{(1)} \right) \,,
    \label{eq:beam_continuity} \\
    &\partial_\tau
	\begin{pmatrix}
   		 \gamma_b^3 v_{bs,x}^{(1)} \\
  		 \gamma_b v_{bs,y}^{(1)}
   	\end{pmatrix}	
    =  -\frac{q_s}{m_s}
	\begin{pmatrix}
   	    \partial_\xi \Psi^{(1)} \\
  		- \partial_y \Psi^{(1)}
   	\end{pmatrix}
    \,,
    \label{eq:beam_momentum}
\end{align}
where $\Psi = A_x- \phi$ is the pseudo-potential encapsulating the electrostatic and electromagnetic field components of the plasma wave~\cite{Chen_IEEE_1987, Lindstrom_arXiV_2025}. Taking the $\tau$-derivative of Eq.~\eqref{eq:beam_continuity}  and substituting the momentum equation yields
\begin{equation}
    \partial^2_\tau n_{bs}^{(1)} + \frac{q_s n_{bs}^{(0)}}{m_s \gamma_b} \left( \gamma_b^{-2} \partial^2_\xi+ \partial^2_y \right) \Psi^{(1)} = 0 \,.
    \label{eq:beam_second_eq_motion}
\end{equation}

To relate $\Psi^{(1)}$ to the perturbed fluid quantities, we use the potential form of the Maxwell-Amp\`ere equation in the Coulomb gauge. Under the QSA, we obtain
\begin{equation}
    \partial_y^2 A_x^{(1)} + \partial_\xi^2\phi^{(1)} = -\rho_b^{(1)} + v_{p,x}^{(1)} \,.
\end{equation}
Integrating the plasma momentum equation \eqref{eq:plasma_eq_motion} gives $v_{p,x}^{(1)} = \Psi^{(1)}$, allowing us to recast the above equation as
\begin{equation}
    \left(\partial_y^2 -1\right) A_x^{(1)} + \left( \partial_\xi^2+1 \right) \phi^{(1)} = -\rho_b^{(1)} \,.
\end{equation}
We can simplify this equation further by noting that
\begin{align}
    \partial_\xi^2 \phi^{(1)} &= (\nabla^2 - \partial_y^2)\phi^{(1)} \nonumber \\
    &= n_p^{(1)}-\rho_b^{(1)} - \partial_y^2 \phi^{(1)} \,.
\end{align}
Substituting this into the previous expression yields the governing equation for the pseudo-potential:
\begin{equation}
    \left( \partial_y^2 - 1\right) \Psi^{(1)} = -n_p^{(1)} \,.
    \label{eq:poisson_QSA}
\end{equation}
Finally, by applying the operator $(\partial_y^2-1)\partial_\tau^2$ to Eq.~\eqref{eq:plasma_oscillations_QSA} and combining the result with Eqs.~\eqref{eq:beam_second_eq_motion} and \eqref{eq:poisson_QSA}, we arrive at the master equation:
\begin{equation}
    \left[ \left( \partial_y^2 - 1 \right) \left( \partial_\xi^2 + 1 \right)\partial_\tau^2 -
   \sum_s \omega_{bs}^2 \left( \partial^2_y + \gamma_b^{-2} \partial^2_\xi \right) \right] n_p^{(1)} = 0 \,,
    \label{eq:full_master_eq}
\end{equation}
where $\omega_{bs}^2 = q_s^2 n_{bs}^{(0)}/m_s \gamma_b$ represents the (normalized) relativistic plasma frequency of species $s$. 

This equation dictates the spatiotemporal evolution of the beam-plasma system. While derived here for $n_p^{(1)}$, the same form holds for other perturbed quantities such as $\rho_b^{(1)}$ or $\Psi^{(1)}$ \cite{San_Miguel_Claveria_short_paper_2026}. As we will see, it captures the full variety of beam-plasma instabilities in the cold-fluid limit, depending on the beam's transverse scale $\sigma_y$. For wide beams ($k_p \sigma_y \gg 1$), the equation describes TSI, OTSI and CFI, whereas for narrow beams ($k_p \sigma_y \ll 1$), it recovers the essential spatiotemporal evolution of SMI and HI, underscoring their kinship with the other streaming plasma instabilities.

In the following sections, we distinguish between transverse modes (OTSI and CFI), which possess a nonzero transverse wavenumber, and longitudinal modes, which depend solely on the longitudinal coordinate. The latter include TSI for an effectively infinite transverse beam extent, as well as SMI and HI for narrow beams. For each regime, we will extract the dominant scaling laws of the instability dynamics and compare the theoretical predictions with PIC simulations.

\section{Transverse unstable modes}
\label{sec:transverse_modes} 

It is well established that for a dilute, highly relativistic, cold beam, OTSI is the fastest-growing instability in an unbounded geometry \cite{Bret_POP_2010}. This instability drives electron plasma waves oscillating at $\omega \sim 1$ (normalized to $\omega_p$) and growing at the maximum growth rate \cite{Bludman_POF_1960, Thode_POF_1976}
\begin{equation}
    \Gamma_{\rm OTSI} = \frac{\sqrt{3}}{2^{4/3}} \left(\sum_s \omega_{bs}^2\right)^{1/3} \left( \frac{k_y^2 + \gamma_b^{-2}}{k_y^2+1} \right)^{1/3} \,,
    \label{eq:OTSI_temporal}
\end{equation}
where $k_y$ is the transverse wavenumber and the longitudinal wavenumber fulfills $k_x v_b \sim k_x \sim 1$. This instability can be viewed as the 2D variant of the purely electrostatic, longitudinal TSI, because waves propagating at oblique angles couple more efficiently with relativistic beam particles due to their anisotropic mass ($\propto \gamma_b^3$ along $x$, $\propto \gamma_b$ along $y$) relative to a disturbance [see Eq.~\eqref{eq:beam_momentum_1}]. Consequently, OTSI is well described in the electrostatic limit \cite{Godfrey_POF_1975, Bret_PRE_2010}, although it retains a weak magnetic component \cite{Bret_PRE_2010, Lemoine_MNRAS_2010}.

Conversely, electromagnetic effects prevail at large angles with $k_x \ll 1$, leading to purely growing, mostly magnetic CFI modes when $k_x=0$. Their maximum growth rate in an unbounded geometry is \cite{Lee_PRL_1973, Godfrey_POF_1975}:
\begin{equation}
    \Gamma_{\rm CFI} = \left(\sum_s \omega_{bs}^2 \frac{k_y^2}{k_y^2+1} \right)^{1/2} \,.
     \label{eq:CFI_temporal}
\end{equation}
A comparison of Eqs.~\eqref{eq:OTSI_temporal} and \eqref{eq:CFI_temporal} shows that the ratio of the CFI to the OTSI growth rates is $\sqrt{8/3}(\sum_s \omega_{bs}^2/2)^{1/6}$ for $k_y \gg 1$. Hence, in the early linear phase, CFI can only rival OTSI for relatively dense, mildly relativistic beams \cite{Godfrey_POF_1975, Bret_POP_2010}. This well-known result, however, applies strictly to unbounded systems in the cold limit.

In this section, we revisit this competition in bounded systems, in the realistic case where the instabilities originate from an initial disturbance at the front of the beam. For this purpose, we assume a perturbation of the form $n_p^{(1)}(\tau,\xi,y) = \delta n_p(\tau,\xi) e^{ik_y y}$ in Eq.~\eqref{eq:full_master_eq}. As we focus on ultrarelativistic beams, we can safely neglect the term $\gamma_b^{-2}\partial_\xi^2$, leading to
\begin{equation}
    \left[ \left( \partial_\xi^2 +1 \right)\partial_\tau^2 - \Gamma_{\rm CFI}^2 \right] \delta n_p = 0 \,.
    \label{eq:CFI_st_master}
\end{equation}

Unlike the purely electrostatic OTSI model of Ref.~\cite{San_Miguel_Claveria_PRR_2022} and the purely inductive CFI model of Ref.~\cite{Pathak_NJP_2015}, Eq.~\eqref{eq:CFI_st_master} captures both electromagnetic and electrostatic effects without relying on the SVEA. Notably, the crossed derivative term $\partial_\xi \partial_\tau$ in Eq.~\eqref{eq:CFI_st_master} does not vanish for large $\gamma_b$. Since this term accounts for spatiotemporal effects, we anticipate that the resulting CFI dynamics will deviate from previous predictions \cite{Pathak_NJP_2015} in the relativistic regime.

\subsection{Qualitative discussion of Eq.~\eqref{eq:CFI_st_master}}
\label{sec:qualitative_discussion}

Before carrying out a comprehensive asymptotic analysis of Eq.~\eqref{eq:CFI_st_master}, we can already identify several distinct behaviors for the solutions, corresponding to different transverse instabilities. The first regime occurs when $\partial_\xi^2 \gg 1$ ($\partial_\xi^2 \gg k_p^2$ in physical units), e.g. near a sharp beam front. In this limit, the equation reduces to
\begin{equation}
    \left( \partial_\tau^2 \partial_\xi^2 - \Gamma_{\rm CFI}^2 \right) \delta n_p = \delta(\xi) \delta(\tau) \,,
    \label{eq:CFI_st_master_front}
\end{equation}
where we have added a Dirac delta source at the beam front to describe the impulse response of the system. This equation can be solved exactly (see Appendix~\ref{sec:AppendixA}): the solution grows asymptotically as $\delta n_p \propto \exp(2\sqrt{\Gamma_{\rm CFI}\tau \xi})$, which represents the spatiotemporal variant of CFI. The validity condition $\partial_\xi^2 \gg 1$ translates into $\xi/\tau \ll \Gamma_{\rm CFI}$. This spatiotemporal solution contrasts with that previously obtained using the SVEA in the fully inductive limit~\cite{Pathak_NJP_2015}:
\begin{equation}
   \delta n_p (\tau, \xi) \sim e^{2\Gamma_{\rm CFI} \sqrt{\xi \tau/Q}} \,,
   \label{eq:CFI_ST_short}
\end{equation}
where $Q = 2\omega_b^2/\gamma_b^2 (k_y^2+1)$. In this solution, the beam plasma frequency $\omega_b$ cancels out in the exponential, leaving a dependence only on $\gamma_b$ and $k_y$. Furthermore, this result was derived under the condition $\xi/\tau \ll Q \sim n_b/\gamma_b^3$ (for $k_y \sim 1$), which is far more restrictive than the condition $\xi/\tau \ll \sqrt{n_b/\gamma_b}$ associated with Eq.~\eqref{eq:CFI_ST_short}.

The other distinct behavior of Eq.~\eqref{eq:CFI_st_master} is found in the limit $\partial_\xi^2 \sim 1$. In this case, substituting $\delta n_p = \hat{n}_p e^{-i \xi}$ into Eq.~\eqref{eq:CFI_st_master} gives
\begin{equation}
    \left[ \partial_\tau^2 \left(\partial_\xi^2 - 2i\partial_\xi \right) - \Gamma_{\rm CFI}^2 \right] \hat{n}_p = \delta(\xi) \delta(\tau) \,.
    \label{eq:OTSI_before_SVEA}
\end{equation}
By applying the SVEA ($\partial_\xi \hat{n}_p \ll \hat{n}_p$) and noting that $\Gamma_{\rm CFI}^2 = \frac{16}{3^{3/2}} \Gamma_{\rm OTSI}^3$, we recover the governing equation for spatiotemporal OTSI under the QSA~\cite{San_Miguel_Claveria_PRR_2022}:
\begin{equation}
    \left( \partial_\tau^2 \partial_\xi - \frac{8i}{3^{3/2}} \Gamma_{\rm OTSI}^3 \right) \hat{n}_p = \delta(\xi) \delta(\tau) \,.
    \label{eq:OTSI_before_SVEA}
\end{equation}
The envelope of this impulsive solution then grows as $\vert \hat{n}_p \vert  \propto e^{\frac{3\sqrt{3}}{4}(\Gamma_{\rm CFI} \tau)^{2/3} \xi^{1/3}} = e^{\frac{3}{2^{2/3}} \Gamma_{\rm OTSI} \tau^{2/3} \xi^{1/3}}$.

The last distinct behavior of Eq.~\eqref{eq:CFI_st_master} is characterized by $\partial_\xi^2 \ll 1$, in which case the solution exhibits purely temporal exponential growth, $\delta n_p \propto \exp(\Gamma_{\rm CFI} \tau)$. This standard CFI solution is valid provided the initial disturbance extends throughout the beam. Yet, as discussed below, this solution is outrun by spatiotemporal OTSI in the $\xi/\tau \gg \Gamma_{\rm CFI}$ region, and therefore does not manifest in the beam-plasma system. 

Finally, we note that purely temporal OTSI, which is expected to arise at $\xi \simeq \tau/3$ in the case of a uniform initial disturbance \cite{San_Miguel_Claveria_PRR_2022}, is not a valid solution of Eq.~\eqref{eq:CFI_st_master}. This shortcoming is a direct consequence of the quasistatic approximation ($\partial_\tau \ll \partial_\xi$). Thus, we will restrict our analysis to $\xi \lesssim \tau/3$, that is, to a limited spatial region extending from the beam front. In practice, this restriction is not overly constraining for the weakly unstable ($\Gamma_{\rm CFI} \ll 1$) beams considered in our study.

In the next section and Appendices \ref{sec:AppendixB}--\ref{sec:AppendixC}, we derive the asymptotic analytical solutions to Eq.~\eqref{eq:CFI_st_master} in greater detail to unravel the competition between CFI and OTSI in flat-top beams. We then validate these results using PIC simulations, which we also extend to Gaussian profiles.

\subsection{Full analytical solution: interplay of spatiotemporal OTSI and CFI}
\label{sec:analytical_eq}

Approximate analytical solutions of Eq.~\eqref{eq:CFI_st_master} are obtained using Laplace transforms combined with the method of steepest descent ~\cite{Swanson_Book_2003, Oughstun_Book_2009}. The double Laplace transform of $\delta n_p$ is defined by
\begin{equation}
    \delta n_p(\alpha,\beta) = \int_0^\infty d\tau \int_0^\infty d\xi\,\delta n_p(\tau,\xi)e^{-i\alpha \tau -i\beta \xi} \,. 
\end{equation}
To derive the system's impulse response (equivalent to its Green's function), we introduce a source term $\delta(\tau) \delta(\xi)$ into the right-hand side of Eq.~\eqref{eq:CFI_st_master} and apply the above transformation, which yields
\begin{equation}
    \delta n_p(\alpha,\beta) = \frac{1}{\alpha^2\beta^2 - \alpha^2 - \Gamma_{\rm CFI}^2} \,.
\end{equation}
Inverting the transform on $\beta$ gives
\begin{equation}
    \delta n_p(\alpha,\xi) =  \frac{1}{2\pi} \int_{\mathcal{C}_\beta} d\beta\, \frac{e^{i\beta \xi}}{\alpha^2 \beta^2 - \alpha^2 - \Gamma_{\rm CFI}^2} \,,
\end{equation}
where the Bromwich contour $\mathcal{C}_\beta$ runs from $\Re \beta = -\infty$ to $\Re \beta = +\infty$, passing below the integrand singularities at $\beta = \pm \sqrt{\Gamma_{\rm CFI}^2/\alpha^2+1}$.

For $\xi <0$, one can close the contour with a semicircle of infinite radius in the half-plane $\Im \beta < 0$, along which the exponential vanishes. Since no singularity is enclosed, the integral is zero.

For $\xi > 0$, applying the residue theorem results in
\begin{align}
    \delta n_p(\alpha, \xi) 
    &= \frac{i H(\xi)}{2 \alpha^2 \sqrt{\frac{\Gamma_{\rm CFI}^2}{\alpha^2}+1}} \nonumber \\
    &\times \left( e^{i\xi \sqrt{\Gamma_{\rm CFI}^2/\alpha^2+1}} - e^{-i\xi \sqrt{\Gamma_{\rm CFI}^2/\alpha^2+1}} \right) \,.
    \label{eq:n_alpha_xi}
\end{align}
where $H(\xi)$ is the unit step function.

To evaluate the time-asymptotic ($\Gamma_{\rm CFI} \tau > 1$) behavior of the inverse transform over $\alpha$, we employ the method of steepest descent. The saddle points of the arguments of the exponentials are determined by
\begin{equation}
    \alpha^2 \sqrt{\Gamma_{\rm CFI}^2+\alpha^2} = \pm \Gamma_{\rm CFI}^2 \xi/\tau \,,
\end{equation}
Squaring this expression leads to a third-order polynomial on $\alpha^2$ that can be exactly solved. We find that all roots are real when $\xi < \xi_{\rm tr}(\tau) \equiv (2/\sqrt{27})\Gamma_{\rm CFI} \tau \simeq 0.385\,\Gamma_{\rm CFI} \tau$. Consequently, we analyze the inverse Laplace transform in three cases: $\xi \ll \xi_{\rm tr}$, $\xi \gg \xi_{\rm tr}$ and $\xi = \xi_{\rm tr}$. A detailed derivation of the corresponding asymptotic solutions, along with a comparison to numerical solutions of the inverse Laplace transform, is provided in Appendix~\ref{sec:AppendixB}.

In the front region ($\xi \ll \xi_{\rm tr}$), the asymptotic solution is
\begin{align}
    \delta n_p (\tau, \xi) &\simeq \frac{H(\xi) H(\tau)}{4\sqrt{\pi}\Gamma_{\rm CFI} (\Gamma_{\rm CFI} \tau \xi)^{1/4}} \left(1 + \frac{3\xi}{8\Gamma_{\rm CFI} \tau} \right)\nonumber \\
    &\times e^{2 \sqrt{\Gamma_{\rm CFI} \tau \xi}(1-\xi/4\Gamma_{\rm CFI} \tau)}  \,.
    \label{eq:CFI_ST}
\end{align}
The absence of $\xi$-oscillatory terms indicates that this solution represents the spatiotemporal evolution of CFI modes. This is corroborated by PIC simulations (Sec.~\ref{sec:PIC_sims} and Ref.~\cite{San_Miguel_Claveria_short_paper_2026}), which show that the unstable modes in this region are magnetically dominated. As discussed above, this novel solution for spatiotemporal CFI starkly differs from that obtained using the SVEA in the fully inductive limit~\cite{Pathak_NJP_2015}, and provides an accurate description of the dominant mode of ultrarelativistic beam-plasma instabilities near the beam front.

Deeper into the beam ($\xi \gg \xi_{\rm tr}$), the modulation amplitude evolves as
\begin{widetext}
\begin{align}
    \delta n_p(\tau,\xi) \simeq \frac{H(\xi) H(\tau)}{4 \Gamma_{\rm CFI}} \left(\frac{2}{3\pi \xi}\right)^{1/2} \left\{\left[1 + \frac{1}{6}\left(\frac{\Gamma_{\rm CFI} \tau}{\xi} \right)^{2/3} \left(\frac{1}{4} + i\frac{\sqrt{3}}{2} \right) \right] e^{i \xi + \frac{3}{4}(\sqrt{3}-i)(\Gamma_{\rm CFI} \tau)^{2/3} \xi^{1/3} - i\pi/4} 
    + c.c. \right\} \,.
    \label{eq:OTSI_ST}
\end{align}
\end{widetext}
The presence of longitudinal modulations identifies this solution as the spatiotemporal variant of OTSI. The leading exponential term coincides exactly with the result derived under the SVEA in Ref.~\cite{San_Miguel_Claveria_PRR_2022} for purely electrostatic oblique modes.

Finally, for an observer moving away from the beam front at the transition boundary $\xi = \xi_{\rm tr}$, the solution reduces to
\begin{equation}
    \delta n_p(\tau, \xi_{\rm tr}) \simeq \frac{3^{1/6} \Gamma(1/3)}{4\sqrt{2}\pi \Gamma_{\rm CFI} (\Gamma_{\rm CFI} \tau)^{1/3}} e^{\frac{4}{3} \sqrt{\frac{2}{3}} \Gamma_{\rm CFI} \tau} \,,
\label{eq:CFI_T}
\end{equation}
where $\Gamma(1/3) \simeq 2.6789$ denotes the gamma function. An observer at this location sees essentially exponential growth at a temporal rate of $(4/3)\sqrt{2/3} \Gamma_{\rm CFI} \simeq 1.09 \Gamma_{\rm CFI}$.

These asymptotic solutions provide a unified account of CFI and OTSI driven by a semi-infinite beam: spatiotemporal CFI governs the front region, whereas spatiotemporal OTSI dominates further behind, with the transition occurring at $\xi = \xi_{\rm tr} \simeq 0.385 \Gamma_{\rm CFI}\tau$. This behavior characterizes the system's response to a pulse disturbance at the beam front, a configuration expected to be relevant when a beam enters a sharp-boundary plasma. For a beam directly generated within the plasma, the instabilities would instead likely grow from a disturbance extending initially throughout the beam. This situation, treated in Appendix~\ref{sec:AppendixC}, allows CFI to grow in a purely temporal manner in the $\xi >\xi_{\rm tr}$ region. Yet it remains superseded there by spatiotemporal OTSI, meaning the results from the impulsive solution remain valid up to $\xi \simeq \tau/3$. Further behind the front, a uniform initial noise would trigger temporal OTSI~\cite{San_Miguel_Claveria_PRR_2022}, which, as previously noted, the present model does not capture.  

\begin{figure}
    \centering
    \includegraphics[width=1\linewidth]{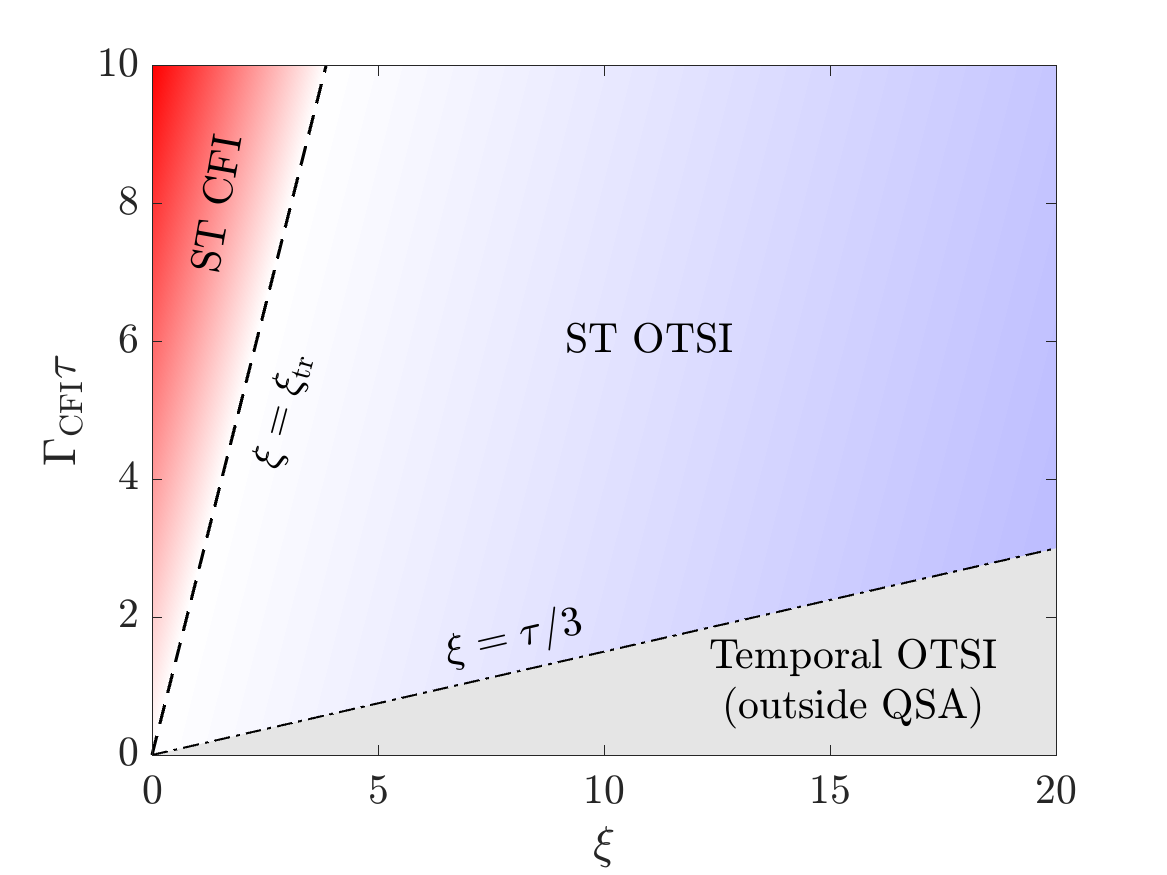}
    \caption{Map of dominant instabilities in $\xi$--$\tau$ space, as derived from Eq.~\eqref{eq:CFI_st_master} and the OTSI theory of Ref.~\cite{San_Miguel_Claveria_PRR_2022} for a semi-infinite uniform beam with the front at $\xi = 0$. The blue area is dominated by spatiotemporal OTSI [Eq.~\eqref{eq:OTSI_ST}], while the gray area is dominated by purely temporal OTSI assuming a spatially extended instability seed~\cite{San_Miguel_Claveria_PRR_2022}. The red area is governed by spatiotemporal CFI [Eq.~\eqref{eq:CFI_ST}]. See text for a detailed discussion of the transition boundaries.
    } 
    \label{fig_long:fig1}
\end{figure}

The regions of dominance for these different instability regimes in $\xi$--$\Gamma_{\rm CFI}\tau$ space are depicted in Fig.~\ref{fig_long:fig1}.
A beam slice at $\xi > 0$ initially undergoes temporal OTSI, which grows at the rate $\Gamma_{\rm OTSI}$ (grey area) \cite{San_Miguel_Claveria_PRR_2022}. By $\tau \simeq 3\xi$ (dot-dashed line), spatiotemporal OTSI takes over (blue area), which evolves as dictated by Eq.~\eqref{eq:OTSI_ST}. As $\Gamma_{\rm CFI}$ decreases, this transition shifts to earlier normalized times $\Gamma_{\rm CFI} \tau$, causing the boundary between the two regimes to asymptotically approach the horizontal axis.

The red area in Fig.~\ref{fig_long:fig1} represents the beam front region dominated by spatiotemporal CFI, which grows according to Eq.~\eqref{eq:CFI_ST}. The dashed line marks the transition between the spatiotemporal CFI and OTSI regimes at $\xi = \xi_{\rm tr}$. In $\xi$--$\Gamma_{\rm CFI}\tau$ space, this boundary is independent of beam-plasma parameters. Consequently, at the time of spatiotemporal CFI saturation, which occurs after $N_e$ temporal $e$-folds at $\xi=\xi_{\rm tr}$, the CFI modes extend over $\xi \sim 0.4 N_e$. For a representative $N_e \simeq 10$, the magnetic filaments occupy a region extending $\sim 4$ plasma skin depths from the beam front, a result that holds regardless of the (small) beam density ratio or (large) Lorentz factor.

\begin{figure}
    \centering
    \includegraphics[width=1\linewidth]{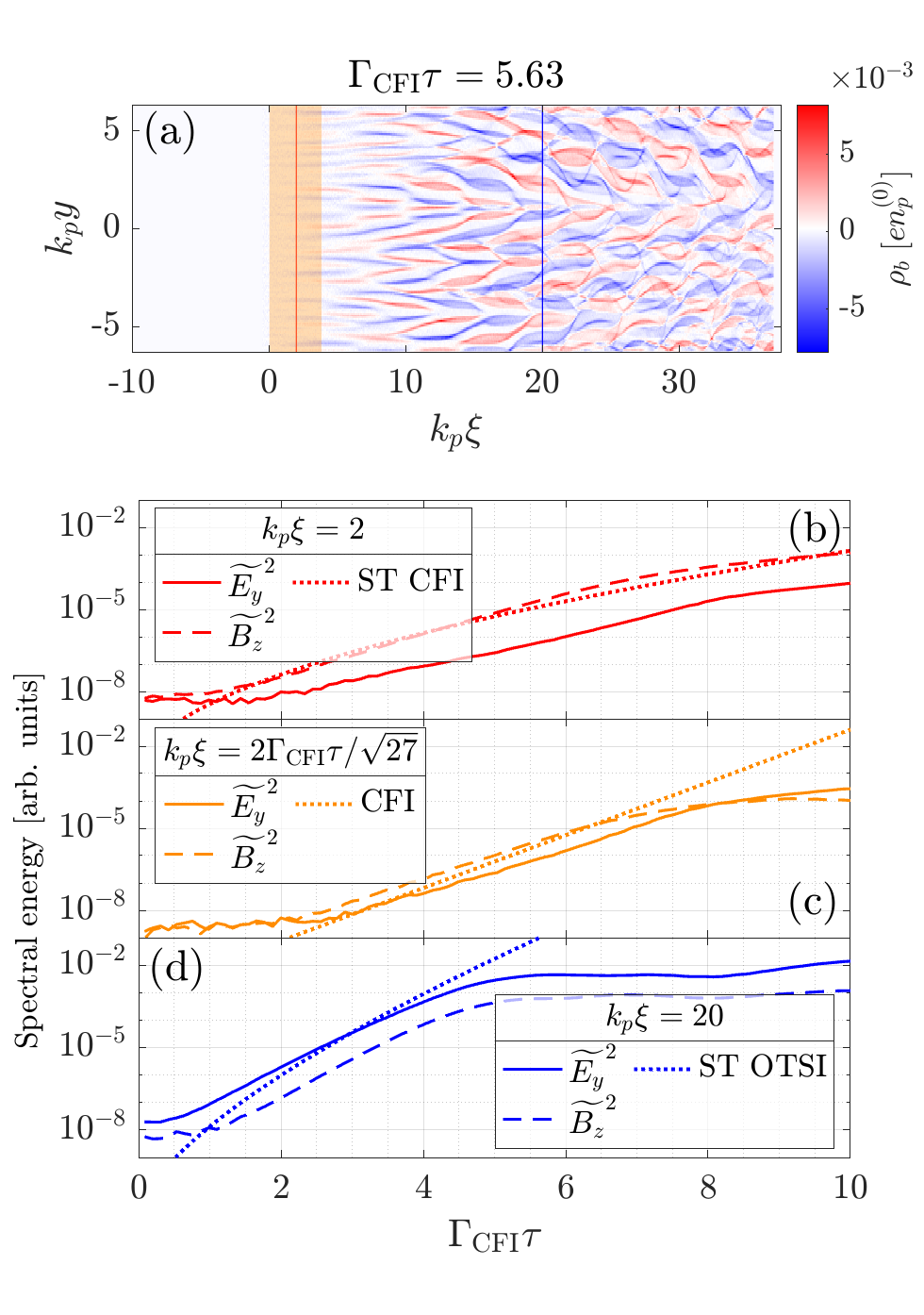}
    \caption{2D PIC pair-beam simulation with $\gamma_b = 2\times 10^4$ and a uniform density $n_b = 0.02n_p$ for both beam electrons and positrons. (a) Snapshot of the beam charge density after a propagation time of $\Gamma_{\rm CFI}\tau = 5.63$. (b)-(d) Temporal evolution of the spectral energy of the $E_y$ (solid lines) and $B_z$ (dashed lines) fields at different longitudinal positions in the beam: (b) $k_p \xi = 2$ [red line in (a)], (c) $k_p \xi = k_p\xi_{\rm tr} = 2\Gamma_{\rm CFI}\tau /\sqrt{27}$ [orange area in (a)] and (d) $k_p \xi = 20$ [blue line in (a)]. Dotted lines represent theoretical predictions for (b) spatiotemporal CFI, (c) temporal CFI (with growth rate $1.09\Gamma_{\rm CFI}$), and (d) spatiotemporal OTSI. These predictions assume $k_y \simeq 3\,k_p$ as measured in the simulation.
    }
    \label{fig_long:fig2}
\end{figure}

\subsection{PIC simulation}
\label{sec:PIC_sims}

We now validate the preceding analytical solutions against a 2D3V (two-dimensional in space, three-dimensional in momentum space) PIC simulation performed with the \textsc{calder} code~\cite{Lefebvre_NF_2003}. Note that hereafter we revert to physical units when discussing simulation results. We consider a charge-neutral, electron-positron pair beam with $\gamma_b = 2 \times 10^4$ and a semi-infinite, flat-top density profile where $n_b = 0.02\,n_p$ for both electrons and positrons. The plasma electrons and ions (protons) are initialized at a temperature of $10\,\rm eV$. The simulation employs a moving window to follow the beam propagation. The domain extends over $-10 \le k_p\xi \le 40$ longitudinally and $-6 \le k_p y \le 6$ transversely, discretized into $1250 \times 150$ cells with 10 macroparticles per cell for each (beam and plasma) species. The time step is $\omega_p \Delta t=0.037$. Boundary conditions are absorbing along $x$ and periodic along $y$ for both fields and particles. The unstable modes develop from the particle-discreteness noise inherent to the PIC method. 

Figure~\ref{fig_long:fig2}(a) shows the beam density distribution after a propagation time of $\Gamma_{\rm CFI}\tau = 5.63$. The image reveals the predicted spatial hierarchy: perpendicular modulations characteristic of CFI are dominant near the beam front ($k_p\xi \lesssim 5$), while oblique modulations typical of OTSI emerge further downstream ($k_p \xi \gtrsim 5$).

The temporal evolution of the spectral energy for the transverse electric field ($\widetilde{E_y}^2$) and the out-of-plane magnetic field ($\widetilde{B_z}^2$) is presented in Figs.~\ref{fig_long:fig2}(b)--(d) for three representative positions: $k_p \xi = 2$, $k_p \xi = k_p \xi_{\rm tr}$, and $k_p \xi = 20$. Spatial Fourier spectra are evaluated using a longitudinal window of $\pm k_p^{-1}$ width at each position. To isolate the contributions from OTSI and CFI modes, the spectral power of $E_y$ is integrated over $k_x/k_p \in (0.5,1.5)$ and $k_y/k_p \in (0,10)$, while that of $B_z$ is integrated over $k_x/k_p \in (-0.5,0.5)$ and $k_y/k_p \in (0,10)$. We note that similar trends are observed when analyzing the root-mean-squared field amplitudes \cite{San_Miguel_Claveria_short_paper_2026}.

At $k_p\xi = 2$ [Fig.~\ref{fig_long:fig2}(b)], the spectral energy of $B_z$ dominates that of $E_y$, and its temporal evolution aligns with the spatiotemporal CFI growth predicted by Eq.~\eqref{eq:CFI_ST}. By contrast, at $k_p \xi = 20$ [Fig.~\ref{fig_long:fig2}(d)], the situation is reversed: the transverse electric field energy prevails, and its growth follows the spatiotemporal OTSI behavior described by the leading exponential term in Eq.~\eqref{eq:OTSI_ST}. Finally, at the transition boundary $\xi \simeq \xi_{\rm tr}(\tau)$, corresponding to $0 \le k_p\xi \lesssim 4$ over the simulation timespan [see orange area in Fig.~\ref{fig_long:fig2}(a)], the magnetic field remains dominant, with a growth rate that matches the predicted purely temporal CFI growth (dotted line).

These simulation results, therefore, support the predicted asymptotic behaviors of spatiotemporal CFI and OTSI in a semi-infinite flat-top beam, as well as their dynamical transition across the beam. In the next section, we extend our analysis to Gaussian beam profiles by numerically solving Eq.~\eqref{eq:CFI_st_master}.

\subsection{Gaussian beams}
\label{sec:gaussian_beams}

Flat-top density distributions, though analytically convenient, are evidently highly idealized, and thus poorly represent the physical profiles generally involved in beam-plasma interactions. This is particularly so for accelerator-based experiments, where beams are more accurately described by Gaussian longitudinal distributions. Moreover, facilities investigating beam-plasma interactions are often designed to maximize peak current, which can result in beam lengths ($\sigma_x$) comparable to the skin depth of standard plasma sources~\cite{Yakimenko_PRAB_2019}.

We now demonstrate that the numerical solution of Eq.~\eqref{eq:CFI_st_master} efficiently captures the spatiotemporal growth of instabilities excited by a short ($\sigma_x \sim k_p^{-1}$), cold relativistic beam with a Gaussian longitudinal density profile. For simplicity, we maintain a uniform transverse beam profile. We note that the case of a transversely Gaussian profile was recently addressed by Walter \textit{et al.}~\cite{Walter_PRE_2024}, who showed that the growth rate varies locally according to the transverse beam distribution.

\begin{figure}
    \centering
    \includegraphics[width=\linewidth]{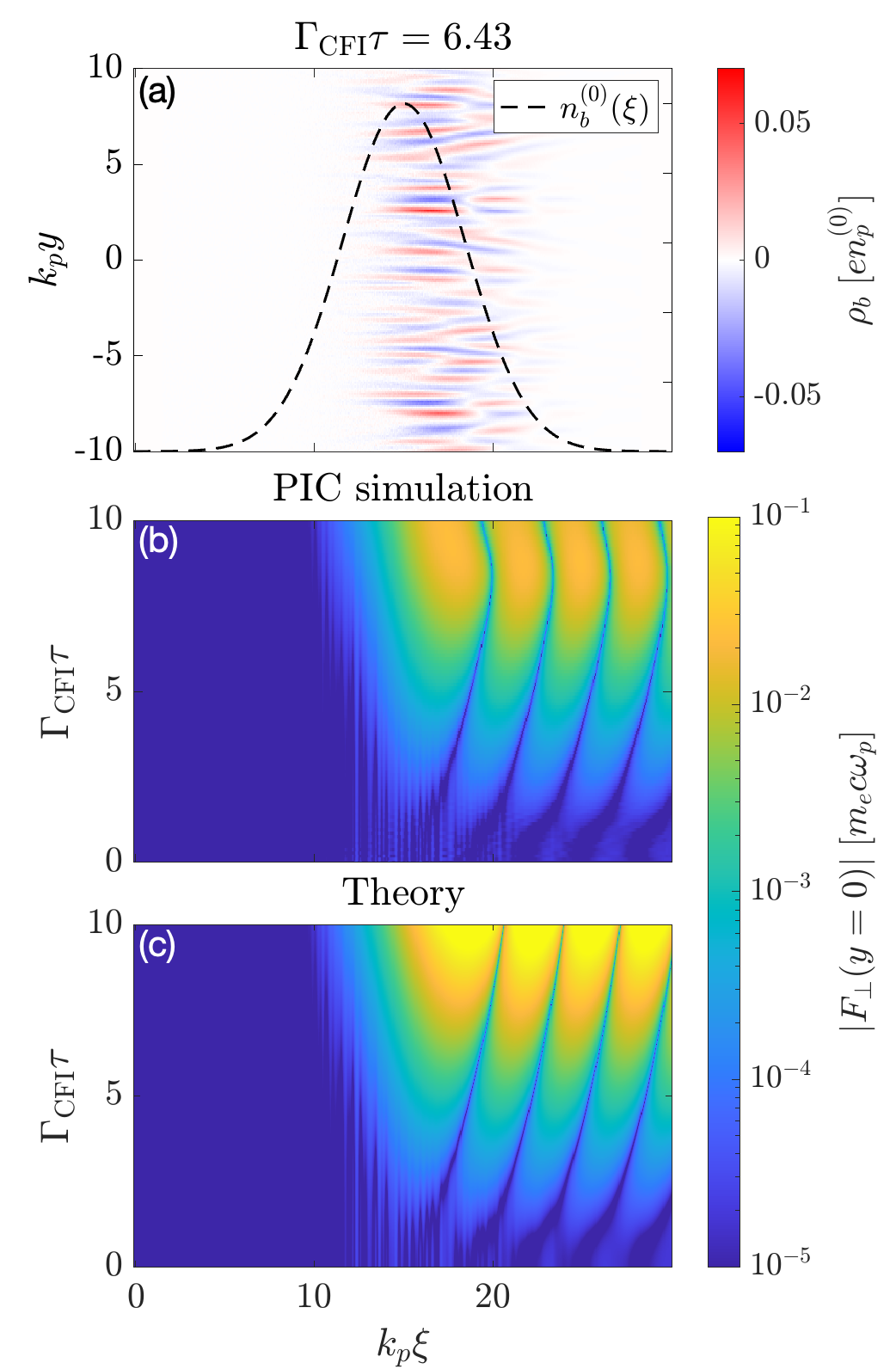}
    \caption{2D PIC pair-beam simulation with Lorentz factor $\gamma_b = 2\times 10^3$ and peak density $n_b/n_p =  0.03$ for both beam electrons and positrons. The longitudinal beam density profile is Gaussian with a root-mean-square length of $k_p\sigma_x = 3.5$ and its center is located at $k_p \xi = 15$. (a) Snapshot of the beam charge  density after a propagation time of $\Gamma_{\rm CFI}\tau = 6.43$. The dashed line shows the initial longitudinal density profile. (b,c) $\xi$--$\tau$ evolution of the norm of the transverse force, $\vert F_\perp (y = 0)\vert$, as (b) extracted from the PIC simulation or (c) computed from Eq.~\eqref{eq:CFI_st_master} with a $\xi$-dependent $\Gamma_{\rm CFI}$ (see text for details).}
    \label{fig:Numerical_1}
\end{figure}

First, we observe that introducing a longitudinal dependence in the unperturbed beam densities $n_{bs}^{(0)}(\xi)$ changes Eq.~\eqref{eq:beam_second_eq_motion} to
\begin{align}
    &\partial^2_\tau n_{bs}^{(1)} + \frac{q_s n_{bs}^{(0)}}{m_s \gamma_b} \left[ \gamma_b^{-2} \partial^2_\xi + \gamma_b^{-2} \left( \partial_\xi \ln n_b^{(0)} \right) \partial_\xi + \partial^2_y \right] \Psi^{(1)} \nonumber \\
    & = 0 \,.
    \label{eq:beam_second_eq_motion_v2}
\end{align}
Combining this equation with Eqs.~\eqref{eq:plasma_oscillations_QSA} and \eqref{eq:poisson_QSA} yields the modified master equation:
\begin{align}
    &\bigg[ \left( \partial_y^2 - 1 \right) \left( \partial_\xi^2 + 1 \right) \partial_\tau^2
    - \sum_s \omega_{bs}^2(\xi) \Big( \partial^2_y + \gamma_b^{-2} \partial^2_\xi \nonumber \\
    &+ \gamma_b^{-2} \partial_\xi \ln n_{bs}^{(0)} \partial_\xi \Big) \bigg]  n_p^{(1)} = 0 \,,
    \label{eq:full_master_eq_v2}
\end{align}
where $\omega_{bs}(\xi) \propto \sqrt{n_{bs}(\xi)}$ is the now $\xi$-dependent plasma frequency of beam species $s$. Assuming again that $\gamma_b \gg 1$, we can discard the two last terms proportional to $\gamma_b^{-2}$. Thus, the governing equation~\eqref{eq:CFI_st_master} remains valid for arbitrary longitudinal beam profiles, provided we account for the spatial dependence of the CFI growth rate as
\begin{equation}
    \Gamma_{\rm CFI}(\xi) = \left( \sum_s \omega_{bs}^2(\xi) \frac{k_y^2}{k_y^2+1} \right)^{1/2} \,.
    \label{eq:CFI_growthrate_v2}
\end{equation}

To validate this result, we conduct a 2D pair beam-plasma simulation using a Gaussian beam with root-mean-square length $k_p \sigma_x = 3.5$ and peak density $n_b = 0.03\,n_p$ for both species. The Lorentz factor is $\gamma_b = 2 \times 10^3$. The moving window spans $0 \le k_p\xi \le 30$ and $-10 \le k_p y \le 10$, with the maximum beam density reached at $k_p\xi_{\rm max} = 15$. All other parameters remain as described in the previous section.

In Fig.~\ref{fig:Numerical_1}, we compare the results of this simulation to the numerical solution of Eq.~\eqref{eq:CFI_st_master} using  $\Gamma_{\rm CFI} (\xi)$ from Eq.~\eqref{eq:CFI_growthrate_v2}, which peaks at $\xi_{\rm max}$ at a value of $\simeq 5.5 \times 10^{-3}\omega_p$ (for the measured transverse wavenumber $k_y \simeq 5\,k_p$). Figure~\ref{fig:Numerical_1}(a) shows the simulated beam charge distribution after a propagation time of $\Gamma_{\rm CFI}\tau = 6.43$. The initial beam profile is overlaid as a dashed line. Parallel filaments form just ahead of the beam density peak ($k_p\xi \simeq 14$), before becoming tilted and longitudinally modulated a few $k_p^{-1}$ away in the decreasing density ramp. This behavior confirms the transition from CFI to OTSI at increasing depths into the beam.

Figure~\ref{fig:Numerical_1}(b) displays the spatiotemporal evolution of the norm of the transverse Lorentz force $F_\perp \equiv e(E_y - cB_z)$ (normalized to $m_e \omega_p c$) as measured along $y = 0$ in the PIC simulation. We use $F_\perp$, rather than $n_b$ or $n_p$, as a figure of merit because the zeroth-order pseudo-potential $\Psi^{(0)}$ vanishes for a neutral pair beam, and $F_\perp \propto \nabla_\perp \Psi$ is also expected to be governed by Eq.~\eqref{eq:CFI_st_master}. Therefore, this quantity can be directly compared to the numerical solution of this equation, which is shown in Fig.~\ref{fig:Numerical_1}(c). For this computation, we take the initial profile $F_\perp(\xi,\tau=0)$ from the PIC simulation and assume a stationary boundary value $F_\perp(\xi=0,\tau>0) = F_\perp(\xi=0,\tau=0)$. The agreement with the simulated linear stage of the instabilities is excellent in both amplitude and phase throughout the simulation window. For $\Gamma_{\rm CFI} \tau \gtrsim 7$ and $k_p\xi \gtrsim 15$, the phase velocity of the simulated OTSI modulations changes sign in $\xi$--$\tau$ space, signaling nonlinear saturation effects not captured by our perturbative model.

\begin{figure}
    \centering
    \includegraphics[width=0.45\textwidth]{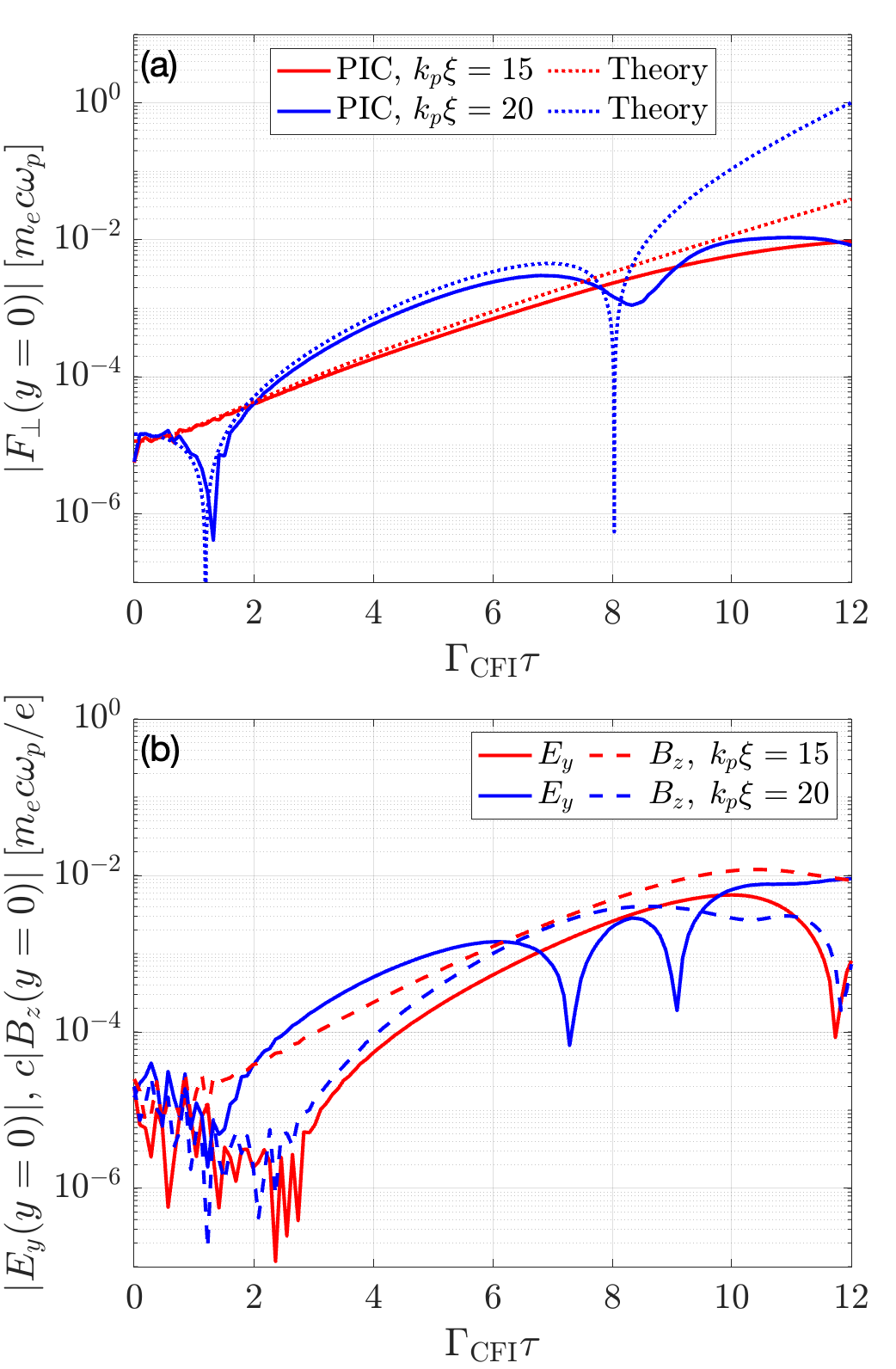}
    \caption{(a) Temporal evolution of $\vert F_y(y=0) \vert$ at longitudinal positions $k_p \xi = 15$ (red) and $k_p \xi = 20$ (blue). Solid lines are from the PIC simulation in Fig.~\ref{fig:Numerical_1} while dotted lines represent the numerical solution of Eq.~\eqref{eq:CFI_st_master}. (b) Temporal evolution of $\vert E_y(y=0) \vert$ (solid) and $\vert B_z(y=0) \vert$ (dashed) at the same longitudinal positions as in (a).
    }
    \label{fig:Numerical_2}
\end{figure}

The close match between the PIC simulation and our model is further illustrated in Fig.~\ref{fig:Numerical_2}(a), which plots $F_\perp$ versus $\tau$ at the beam center ($k_p\xi = 15$, red), and at the trailing edge ($k_p\xi = 20$, blue). The solid and dashed lines represent the simulation and theoretical results, respectively. Consistent with Fig.~\ref{fig:Numerical_1}(a), the steady growth (without oscillations) seen at $k_p \xi = 15$ is typical of CFI, while the oscillatory growth seen at $k_p\xi = 20$ is characteristic of OTSI.

Figure~\ref{fig:Numerical_2}(b) complements these results by plotting the evolution of $\vert E_y \vert$ (dashed curves) and $\vert B_z \vert$ (solid curves) versus $\tau$ at the same longitudinal positions. As expected, the growing fields are predominantly magnetic at the beam center. At the beam rear, the hierarchy is less clear: the transverse electric field initially prevails but is overtaken by the magnetic field around $\Gamma_{\rm CFI} \tau \simeq 6$, shortly before a node and the onset of nonlinear effects as discussed above. While the magnetic field saturates around $\Gamma_{\rm CFI} \tau \simeq 8$, the electric field continues to grow with visible oscillations, until it prevails again beyond $\Gamma_{\rm CFI} \tau \simeq 10$.
Since our main Eq.~\eqref{eq:CFI_st_master} does not directly describe $E_y$ or $B_z$, but rather governs $n_p^{(1)}$, $\rho^{(1)}$, or $\Psi^{(1)}$, we do not overlay theoretical curves in Fig.~\ref{fig:Numerical_2}(b). In principle, though, it would be possible to separately derive $E_y^{(1)}$ and $B_z^{(1)}$ from $\rho^{(1)}$ following the method of Ref.~\cite{Keinigs_POF_1987}.

These results illustrate the case of a relatively short beam, with $k_p \sigma_x = \mathcal{O}(1)$. An additional PIC simulation using $k_p\sigma_x \simeq 20$ (not shown here) reveals that $E_y$ eventually becomes dominant for longer bunches.

To summarize this section, we have used the QSA to describe, within a unified electromagnetic framework, the linear growth of instabilities driven by a cold, ultrarelativistic beam with a well-defined leading front and infinite transverse extent. We have focused on the dominant CFI and OTSI modes, both characterized by a finite transverse wavenumber. By relaxing the SVEA, this model remains valid at arbitrary distances from the front of a top-hat beam, or for Gaussian beams with lengths as short as a single plasma skin depth. This result is particularly relevant for accelerator-based experiments, where such beam configurations are common.

\section{Longitudinal modes}
\label{sec:Longitudinal_modes}

We now extend the above framework to purely longitudinal modes. We first address the two-stream instability for a transversely wide beam ($k_p \sigma_y \gg 1$), followed by two instabilities that arise in transversely narrow beams ($k_p \sigma_y \lesssim 1$): the self-modulation instability and the hosing instability.

\subsection{Wide beams: Two-stream instability}

In the relativistic limit, the purely electrostatic TSI governs the beam-plasma interaction only for wide, sufficiently hot beams, either from the onset of the interaction \cite{Rudakov_JETP_1971, Bret_POP_2010} or during a subsequent stage once the fastest-growing OTSI has saturated by heating the beam transversely \cite{Fainberg_JETP_1970, Breizman_JETP_1971, Bret_POP_2010, Sironi_APJ_2014}. Since our model neglects thermal effects, the assumption of a purely one-dimensional oscillation spectrum is most relevant when the beam propagates along a strong longitudinal magnetic field \cite{Fainberg_JETP_1970, Thode_POF_1975a}.

\begin{figure}
    \centering
    \includegraphics[width=\linewidth]{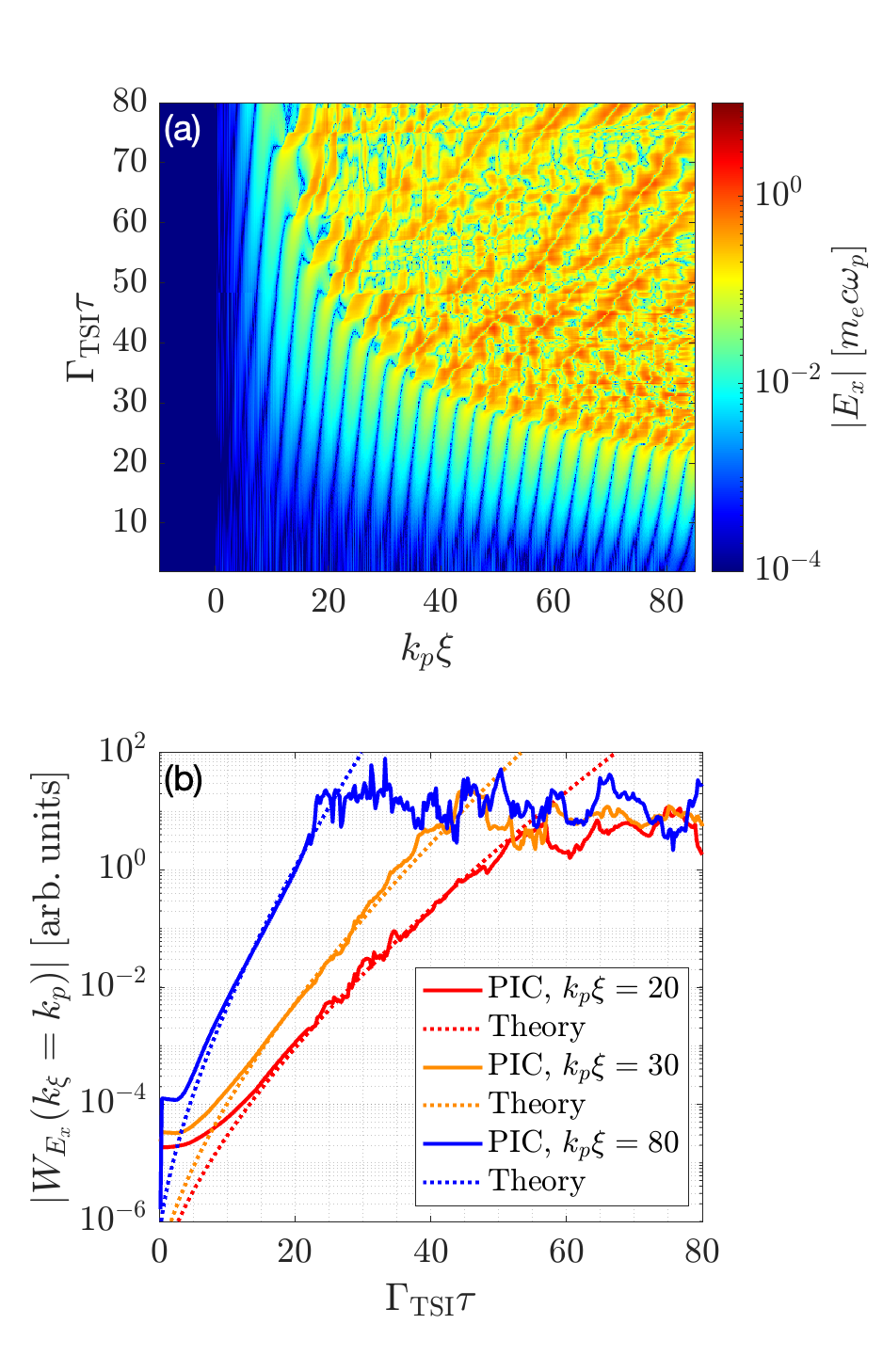}
    \caption{1D PIC simulation of a pair beam-plasma interaction with $\gamma_b = 200$ and $n_b/n_p = 0.6$. (a) Spatiotemporal evolution of the norm of the longitudinal electric field, $\vert E_x \vert$. (b) Temporal evolution of the Wigner-Ville distribution of $E_x$ at different longitudinal positions (solid lines). Dashed lines represent the amplitude of the asymptotic solution given by  Eq.~\eqref{eq:TSI_QSA_analyt_GF}. See text for further details.
    }
    \label{fig:Numerical_TSI}
\end{figure}

Within the quasistatic framework of Sec.~\ref{sec:QSA_model}, setting the transverse derivatives to zero in Eq.~\eqref{eq:full_master_eq} leads to
\begin{equation}
    \left[ \partial_\tau^2 \left( \partial_\xi^2 + 1 \right) +
     \sum_s \frac{\omega_{bs}^2}{\gamma_b^2} \, \partial^2_\xi \right] n_p^{(1)} = 0 \,.
\end{equation}
For purely longitudinal modes, we express the density perturbation as $n_p^{(1)}(\xi,\tau) = \hat{n}_p (\xi,\tau) e^{-i \xi}$, where $\hat{n}_p$ represents the amplitude of the longitudinal modulation. Applying the SVEA, this equation simplifies to 
\begin{equation}
    \left( \partial_\tau^2 \partial_\xi - \frac{8i}{3^{3/2}} \Gamma_{\rm TSI}^3
    \right) \hat{n}_p = 0 \,,
    \label{eq:TSI_QSA}
\end{equation}
where 
\begin{equation}
    \Gamma_{\rm TSI} =\frac{\sqrt{3}}{2^{4/3}} \left(\sum_s \frac{\omega_{bs}^2}{\gamma_b^2} \right)^{1/3}
\end{equation}
is the temporal growth rate of TSI in an unbounded system~\cite{Bludman_POF_1960}. This equation is formally equivalent to that derived for OTSI in the electrostatic limit~\cite{San_Miguel_Claveria_PRR_2022}. It can also be recovered from the $(\omega,k_x)$ dispersion relation by substituting $\omega \to 1+ i \partial_t$ and $k_x \to 1 - i\partial_x$ under the SVEA~\cite{Rostomyan_POP_2000}. Its asymptotic impulse solution is given by~\cite{Evans_POF_1970, Rostomyan_POP_2000}
\begin{equation}
    \hat{n}_p(\tau,\xi) \propto e^{\frac{\sqrt{3}}{2^{2/3}} (\sqrt{3}+i) \Gamma_{\rm TSI} \xi^{1/3} \tau^{2/3}} \,.
    \label{eq:TSI_QSA_analyt_GF}
\end{equation}

Figure~\ref{fig:Numerical_TSI} compares this asymptotic solution with results from a 1D PIC simulation of a pair beam-plasma system. The beam has a top-hat density profile over $0 \le k_p\xi \le 90$, with $n_b/n_p = 0.6$ and $\gamma_b = 200$ for both beam electrons and positrons. The $\xi$--$\tau$ evolution of the norm of the longitudinal electric field ($\vert E_x \vert$) is shown in Fig.~\ref{fig:Numerical_TSI}(a). The data illustrate the transition from the linear regime, characterized by regular oscillations with $k_x \simeq 1$, to the nonlinear regime, marked by strong spectral broadening. Because the local growth rate increases with $\xi$ (in the $\xi \ll \tau$ limit considered here), the instability reaches saturation earlier at larger $\xi$.

To track the evolution of the envelope of the dominant mode, we compute the Wigner-Ville distribution~\cite{Cohen_book_1995} of the spatial profile of $E_x$ every 2000 time steps. Figure~\ref{fig:Numerical_TSI}(b) plots the resulting amplitude, $\vert W_{E_x} \vert$, at $k_x \simeq 1$ for three longitudinal positions (solid lines). The predicted growth of the envelope [the norm of Eq.~\eqref{eq:TSI_QSA_analyt_GF}] is overlaid as dash-dotted lines, showing excellent agreement with the PIC data. A less quantitative comparison between Eq.~\eqref{eq:TSI_QSA_analyt_GF} and a PIC simulation was reported in Ref.~\cite{Jones_POF_1983}.

While TSI is a cornerstone of plasma micro-turbulence studies~\cite{Ewart_PNAS_2025}, its linear evolution in the relativistic regime is typically overtaken by transverse instabilities whenever the beam's transverse dimensions are considered. However, a second family of longitudinal instabilities exists with growth rates comparable to their transverse counterparts; these arise in systems where the beam has a finite transverse size on the order of, or smaller than, the plasma skin depth.

\subsection{Narrow beams: Self-modulation and hosing instabilities}

The self-modulation (SMI,~\cite{Kumar_PRL_2010, Schroeder_PRL_2011}) and hosing (HI,~\cite{Whittum_POF_1993, Krall_POP_1995}) instabilities are primarily studied in the context of plasma accelerators~\cite{Verra_PRL_2022,Verra_POP_2023}. Although both drive longitudinal beam density modulations, they differ from TSI in that the beam width is smaller than (or comparable with) the plasma skin depth, and the underlying particle motion is predominantly perpendicular to the mean beam velocity. In strictly 1D systems, this transverse motion is suppressed, leaving TSI as the sole unstable mode. Consequently, 2D models are required to capture the excitation of SMI and HI by long relativistic beams ($k_p\sigma_x \gg 1$).
We refer to this class of instabilities as narrow-beam instabilities (NBI).

\begin{figure}
    \centering
    \includegraphics[width=\linewidth]{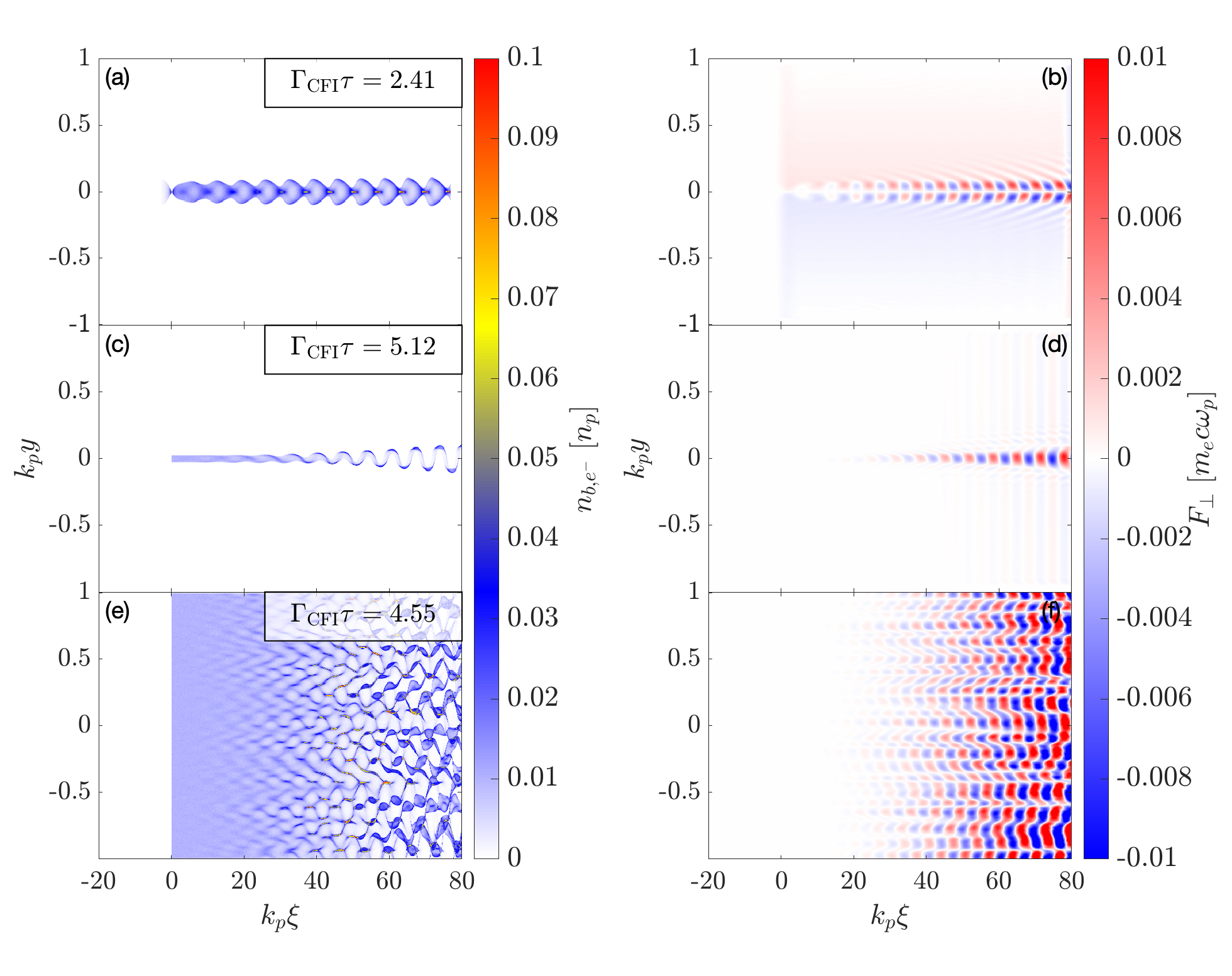}
    \caption{2D PIC simulations of a relativistic particle beam experiencing SMI (a,b), HI (c,d) and OTSI (e,f). SMI is driven by an electron beam, whereas HI and OTSI are driven by a neutral pair beam. Panels (a), (c) and (e) show the beam electron density distribution, while panels (b), (d), and (f) show the associated transverse force $F_\perp = E_y-cB_z$. See text for simulation parameters.
    }
    \label{fig:SMI_HI}
\end{figure}

SMI generates growing modulations in the transverse beam envelope, leading to a train of bunches with $k_x \sim k_p$ [Fig.~\ref{fig:SMI_HI}(a)]. Conversely, HI excites growing beam centroid displacements [Fig.~\ref{fig:SMI_HI}(c)]. Existing analytical models typically track the first and second transverse moments---i.e., the centroid and radius---of the beam distribution~\cite{Schroeder_PRE_2012, Vieira_PRL_2014}. These models often rely on the QSA to derive the equation for the transverse force that closes the fluid system.

Here, we demonstrate that our quasistatic theory can reproduce the linear growth of both instabilities by tracking the unstable electromagnetic mode (the pseudo-potential $\Psi^{(1)}$) and the density perturbations ($n_b^{(1)}$, $n_p^{(1)}$), rather than just the moments. To do so, we consider a 2D beam distribution of the form
\begin{equation}
    n_b^{(0)}(\xi,y) = H(\xi)  \Pi\left( |y|/2\sigma_y \right) \,,
\end{equation}
where $\Pi(x) = H(1-\vert x \vert)$ is the top-hat function.
The assumption of a narrow beam ($k_p \sigma_y \ll 1$) allows us to approximate $\partial_y^2-1 \simeq \partial_y^2$ in the equation for $\Psi^{(1)}$ [Eq.~\eqref{eq:poisson_QSA}]. Physically, this implies that the SMI and HI modes possess a transverse component $k_y \gg 1$. The master equation \eqref{eq:full_master_eq} then becomes
\begin{equation}
    \left[\partial_\tau^2 \left( \partial_\xi^2 + 1 \right) - 
    \sum_s \omega_{bs}^2 \right] n_p^{(1)} = 0 \,,
\end{equation}
Writing $n_p^{(1)}(\xi,\tau) = \hat{n}_p(\xi,\tau)e^{-i \xi + i k_y y}$, and applying the SVEA leads to the following equation for the perturbation envelope:
\begin{equation}
    \left( \partial_\tau^2 \partial_\xi - \frac{8i}{3^{3/2}} \Gamma_{\rm NBI}^3 \right) \hat{n}_p = 0 \,.
    \label{eq:SMI_QSA}
\end{equation}
where
\begin{equation}
    \Gamma_{\rm NBI} \equiv \Gamma_{\rm OTSI}(k_y \to \infty) =  \frac{\sqrt{3}}{2^{4/3}} \left(\sum_s \omega_{bs}^2 \right)^{1/3}  
\end{equation}
is the growth rate of the instability excited by a narrow beam. This differential equation matches that derived in Ref.~\cite{Schroeder_PRL_2011} for both the beam radius (for SMI) and the beam centroid displacement (for HI) within the narrow-beam limit, differing only in the definition of the growth rate by a geometrical factor that tends to unity as $k_p \sigma_y \to 0$. This similarity highlights the close connection between SMI, HI and OTSI in the $k_y\gg 1$ limit.

\begin{figure}
    \centering
    \includegraphics[width=\linewidth]{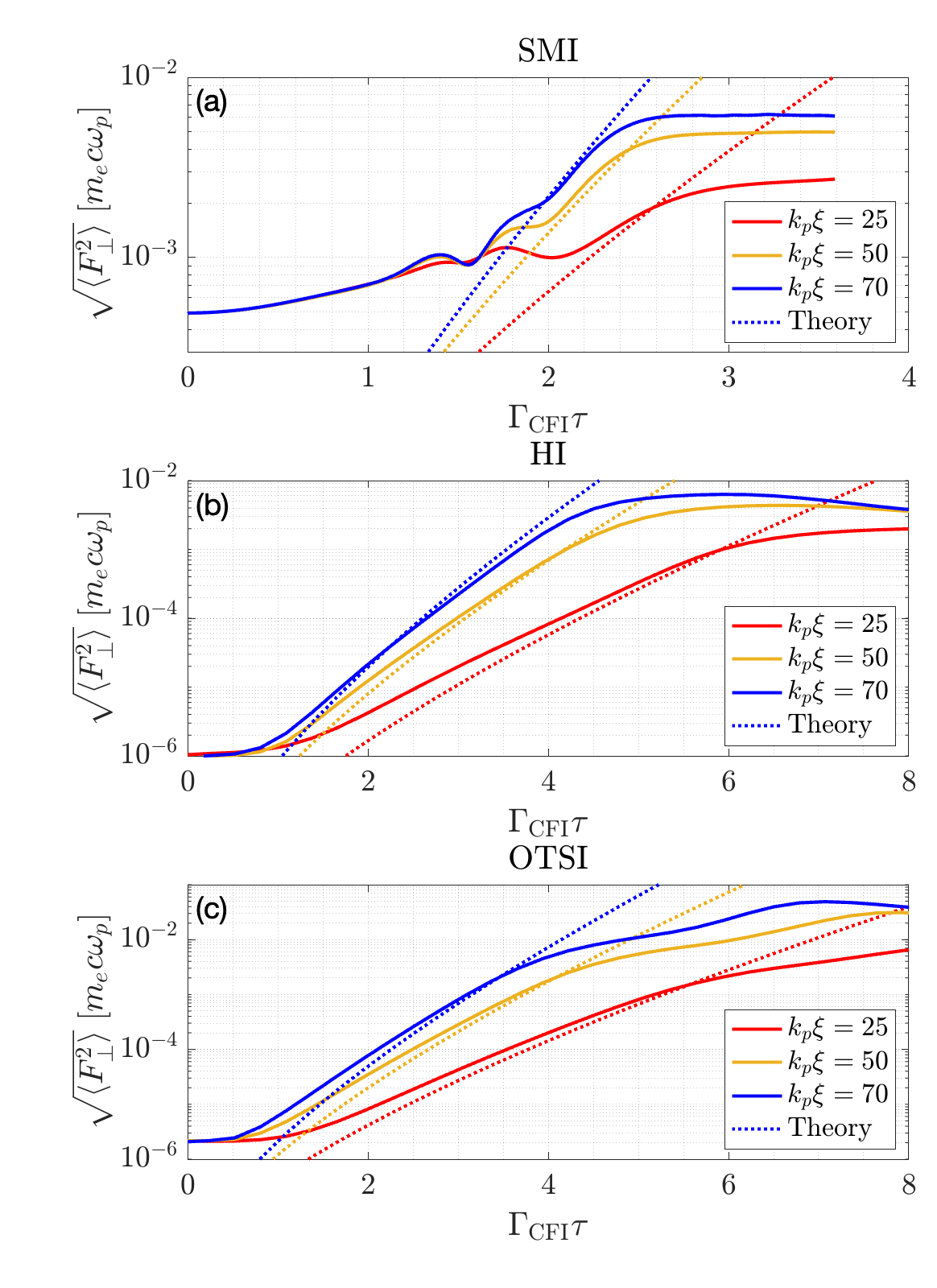}
    \caption{Temporal evolution of the r.m.s. amplitude of the transverse force $F_\perp$, averaged over the initial beam width, at three longitudinal positions. Solid lines are from the (a) SMI, (b) HI and (c) OTSI PIC simulations shown in Fig.~\ref{fig:SMI_HI}; dotted lines represent the asymptotic analytical predictions from (a,b) Eq.~\eqref{eq:SMI_QSA} or (c) Eq.~\eqref{eq:CFI_st_master}.
    }
    \label{fig:SMI_HI_temp}
\end{figure}

As a validation test, we performed 2D PIC simulations of HI and SMI driven by narrow beams. In Figs.~\ref{fig:SMI_HI} and \ref{fig:SMI_HI_temp}, we compare the results to analytical predictions, as well as to an additional OTSI simulation with $k_p \sigma_y = \infty$.

To excite SMI, we used a nonneutral electron beam with $k_p \sigma_x = 80$, $k_p \sigma_y = 0.2$, $n_b/n_p = 0.01$ and $\gamma_b = 2000$. To reduce the impact of the plasma wakefield excited by this nonneutral beam, we included a small linear density ramp of length $2\pi k_p^{-1}$ at the beam front. For HI and OTSI, we considered a neutral flat-top pair beam with $k_p \sigma_x = 80$, $k_p \sigma_y = 0.025$, $n_b/n_p = 0.01$ (for both beam species) and $\gamma_b = 2000$ (HI) or $k_p \sigma_y = \infty$ (OTSI). The moving window had a size of $100\,k_p^{-1} \times 2\,k_p^{-1}$ with a discretization $k_p \Delta x = k_p \Delta y = 0.005$. Both plasma and beam species were represented by 25 macroparticles per cell, initialized at zero temperature. In order to quench numerical Cherenkov-type instabilities, we combined the specifically designed Godfrey-Vay filter~\cite{Godfrey_JCP_2014} with multi-pass compensated binomial filtering~\cite{Vay_JCP_2011}. Maxwell’s equations were solved using the Cole-Karkkainen scheme~\cite{Karkkainen_2006}, which is dispersion-free along the main axis and allows for a large time step, $\Delta t = 0.9 \Delta x/c = 0.0045 \,\omega_p^{-1}$.

Figure~\ref{fig:SMI_HI} displays the modulated electron beam densities and the transverse force $F_\perp = E_y-cB_z$. The striking similarity between HI [Figs.~\ref{fig:SMI_HI}(c,d)] and OTSI [Figs.~\ref{fig:SMI_HI}(e,f)] simulations confirms their close relationship. For SMI [Figs.~\ref{fig:SMI_HI}(a,b)], the initial wakefield seeding raises the noise level and reduces the number of observable $e$-folds; consequently, its evolution is plotted over a shorter propagation time. Despite this difference, similar growth behavior is seen, as confirmed in Fig.~\ref{fig:SMI_HI_temp}.

SMI and HI inherit the same type of transverse fields as OTSI, but with a different symmetry with respect to the beam axis. For HI, the oscillatory transverse field is maximum on axis for HI and takes the form of an oscillatory dipole field (deflecting beam particles up or down). For SMI, the transverse field vanishes on axis and acts as an oscillatory focusing-defocusing field that drives the modulation of the beam envelope. This common origin underscores why these instabilities share similar growth dynamics.

Finally, we quantify the simulated growth in Fig.~\ref{fig:SMI_HI_temp}. Panels (a-c) compare the root-mean-square amplitude of the transverse force (averaged over the initial beam width) at various longitudinal positions ($\xi$) to the asymptotic analytical solutions of Eq.~\eqref{eq:SMI_QSA}. Accounting for the aforementioned wakefield seed by elevating the initial SMI amplitude by two orders of magnitude yields good agreement across all three instabilities. Unlike previous moment-based models, our approach tracks the coupled evolution of fields and densities, providing a unified framework that highlights  the fundamental relationship between these classes of streaming instabilities.

\section{Conclusions}
\label{sec:conclusions}

We have developed a quasistatic, cold-fluid theory to describe the linear growth of streaming plasma instabilities driven by bounded, ultrarelativistic particle beams. This model provides the first unified framework for all electromagnetic unstable modes in the spatiotemporal regime. For beams with transverse extent $\sigma_y \gg k_p^{-1}$, the current filamentation instability is predicted to dominate at the beam front, exhibiting a dynamics distinct from that previously anticipated~\cite{Pathak_NJP_2015}. By bypassing the slowly-varying envelope approximation, our approach captures transverse modes driven by beams with lengths comparable to the plasma skin depth, a feature particularly relevant for plasma accelerators. Further into the beam, the spatiotemporal oblique two-stream instability takes over, with the transition between CFI- and OTSI-dominated regions occurring at $\xi_{\rm tr} \simeq 2\Gamma_{\rm CFI} \tau/\sqrt{27}$. These predictions are supported by excellent agreement with 2D PIC simulations.

Besides transverse modes, we have applied the model to longitudinal instabilities. In a 1D geometry, the model successfully captures the spatiotemporal growth of the purely electrostatic (parallel) two-stream instability. For transversely narrow beams ($\sigma_y \ll k_p^{-1}$), the model describes the linear growth of the self-modulation and hosing instabilities. We demonstrate that SMI and HI share the same spatiotemporal evolution as OTSI in the limit of large transverse wavenumbers, pointing to a common origin for these instabilities.

These results underscore the power of the quasistatic approach. Beyond its analytical utility, this work advocates the use of quasistatic PIC codes for the large-scale numerical modeling of instabilities driven by highly dilute, ultrarelativistic particle beams \cite{Labro_CPC_2026}. Such configurations, common in high-energy astrophysics, can be computationally inaccessible to standard PIC codes due to the vast disparity between beam and plasma dynamical timescales \cite{Sironi_APJ_2014, Perry_MNRAS_2021}. Quasistatic PIC codes are uniquely suited to bridging this gap, offering a robust path forward for modeling such systems \cite{San_Miguel_Claveria_short_paper_2026}.

\begin{acknowledgments}
P.S.M.C. and F.F. would like to thank Jorge Vieira for insightful discussions on the narrow-beam instabilities.
P.S.M.C. received funding from the European Union under HORIZON-WIDERA-2023-TALENTS-02 (PLAXI project, Grant Agreement No.~101180632). The work at IST was supported by the European Research Council (ERC-2021-CoG Grant XPACE No.~101045172). The work at CEA and LOA was supported by the Agence Nationale de la Recherche (ANR) (UnRIP project, Grant No.~ANR-20-CE30-0030). Computational resources were provided by FCT I.P. on the Deucalion and MareNostrum5 platforms through Project 2025.00196.CPCA.A3 and 2025.00204.CPCA.A3, and by GENCI-TGCC on the supercomputer IRENE under Grants No.~A0100510786 and No.~A0190512993.
\end{acknowledgments}

\appendix

\section{Exact impulsive solution of Eq.~\eqref{eq:CFI_st_master_front}}
\label{sec:AppendixA}

A double Laplace transform of Eq.~\eqref{eq:CFI_st_master_front} yields
\begin{equation}
    \delta n_p(\alpha,\beta) = \frac{1}{\alpha^2 \beta^2 - \Gamma^2} \,.
\end{equation}
To ease notation, we write $\Gamma$ in place of $\Gamma_{\rm CFI}$. The inverse Laplace transform over $\beta$ can be readily solved by applying the residue theorem to the pole singularities at $\beta = \pm \Gamma/\alpha$:
\begin{align}
    \delta n_p(\alpha, \xi) &= \frac{1}{2\pi} \int_{C_\beta} d\beta \frac{e^{i \beta \xi}}{\alpha ^2 \beta^2 - \Gamma^2} \nonumber \\
    & = \frac{i}{2 \Gamma \alpha} \left( e^{i \Gamma \xi /\alpha} - e^{- i \Gamma \xi/\alpha} \right) H(\xi) \,.
\end{align}

The solution $\delta n_b(\tau,\xi)$ is given by the inverse Laplace transform
\begin{align}
    \delta n_p(\tau,\xi) &= \frac{i }{4\pi \Gamma} \int_{C_\alpha} d\alpha \frac{e^{i \alpha \tau}}{\alpha} \left( e^{i \Gamma \xi /\alpha} - e^{- i \Gamma \xi/\alpha} \right) H(\xi) \nonumber \\
    &= \frac{i }{4\pi \Gamma} \int_{C_\alpha} \frac{d\alpha}{\alpha} \left( e^{i(\alpha \tau + \Gamma \xi/\alpha)} - e^{i(\alpha \tau - \Gamma \xi/\alpha)} \right) H(\xi)  \nonumber \\
    & = - \frac{1}{2\Gamma} \left\{ \mathrm{Res}\left[\frac{e^{i(\alpha \tau + \Gamma \xi/\alpha)}}{\alpha};0 \right] - \right. \nonumber \\
    & \hspace{0.6cm}\left. -\mathrm{Res}\left[\frac{e^{i(\alpha \tau - \Gamma \xi/\alpha)}}{\alpha};0 \right]\right\} H(\xi) H(\tau) \,.
\end{align}
The residues can be exactly evaluated by making a Laurent expansion of the two terms around $\alpha=0$ \cite{McKinstrie_POP_1996}:
\begin{align}
    \frac{e^{i(\alpha \tau + \Gamma \xi/\alpha)}}{\alpha} &= \sum_{l,m=0}^\infty \frac{i^{l+m}}{l!m!} \frac{\tau^l (\Gamma \xi)^m}{\alpha^{m-l+1}} \,, \\
    \frac{e^{i(\alpha \tau - \Gamma \xi/\alpha)}}{\alpha} &= \sum_{l,m=0}^\infty \frac{(-1)^m i^{l+m}}{l!m!} \frac{\tau^l (\Gamma \xi)^m}{\alpha^{m-l+1}} \,.
\end{align}
The residue is the $\alpha^{-1}$ coefficient in each expansion, obtained by setting $m=l$:
\begin{align}
    \mathrm{Res}\left[\frac{e^{i(\alpha \tau + \Gamma \xi/\alpha)}}{\alpha};0 \right] &= \sum_{l=0}^\infty \frac{(-\Gamma \xi \tau)^l}{(l!)^2} \,, \nonumber \\
    \mathrm{Res}\left[\frac{e^{i(\alpha \tau - \Gamma \xi/\alpha)}}{\alpha};0 \right] &= \sum_{l=0}^\infty \frac{(\Gamma \xi \tau)^l}{(l!)^2} \,.
\end{align}
We now use the Bessel function identities \cite{Abramowitz_1972}:
\begin{align}
    J_n(z) &= (z/2)^n \sum_{n=0}^\infty \frac{(-z^2/4)^l}{l! (l+n)!} \,, \\
    I_n(z) &= (z/2)^n \sum_{n=0}^\infty \frac{(z^2/4)^l}{l! (l+n)!} \,,
\end{align}
to obtain 
\begin{align}
    \mathrm{Res}\left[\frac{e^{i(\alpha \tau + \Gamma \xi/\alpha)}}{\alpha};0 \right] &= J_0 \left(2\sqrt{ \Gamma \xi \tau} \right)  \,, \nonumber \\
    \mathrm{Res}\left[\frac{e^{i(\alpha \tau - \Gamma \xi/\alpha)}}{\alpha};0 \right] &= I_0 \left(2\sqrt{ \Gamma \xi \tau} \right) \,.
\end{align}
There follows the sought-for exact solution:
\begin{align}
    &\delta n_p(\tau, \xi) = \frac{H(\xi) H(\tau) }{2\Gamma} \nonumber \\
    &\times \left[ I_0 \left(2 \sqrt{ \Gamma \xi \tau} \right) - J_0 \left(2\sqrt{ \Gamma \xi \tau} \right) \right] \,.
\end{align}
The asymptotic evolution of $\delta n_p$ is governed by the modified Bessel function. Using $I_0(z) \sim e^z/\sqrt{2\pi z}$ for $z\gg 1$, one obtains
\begin{equation}
    \delta n_p(\tau,\xi) \simeq \frac{1}{4\sqrt{\pi} \Gamma } \frac{e^{2\sqrt{\Gamma \xi \tau}}}{(\Gamma \xi \tau)^{1/4}}  \,,
    \label{eq:sol_strong_qsa_short_asymp}
\end{equation} 
when $\sqrt{\Gamma \xi \tau} \gg 1$.


\section{Asymptotic impulsive solution of Eq.~\eqref{eq:CFI_st_master}}
\label{sec:AppendixB}

We wish to approximate the inverse Laplace of Eq.~\eqref{eq:n_alpha_xi}:
\begin{align}
    \delta n_p (\tau, \xi) &= \frac{iH(\xi)}{4\pi} \int_{\mathcal{C}_\alpha} d\alpha \, g(\alpha) \left(e^{\tau h_+(\alpha; \theta)} - e^{\tau h_-(\alpha;\theta)}\right)  \nonumber \\
    &= \left( I^+ - I^- \right) H(\xi) \,,  
    \label{eq:np_inverse_LT}
\end{align}
where we have introduced $\theta =  \xi/\tau$ and
\begin{align}
    &g(\alpha) = \frac{1}{\alpha \sqrt{\Gamma^2+\alpha^2}} \,, \\
    &h_\pm (\alpha; \theta) = i\left( \alpha \pm \frac{\theta}{\alpha} \sqrt{\Gamma^2+\alpha^2} \right) \,.  \label{eq:h_pm} \\
    &I^\pm = \frac{i}{4\pi} \int_{\mathcal{C}_\alpha} d\alpha\, g(\alpha) e^{\tau h_\pm(\alpha, \theta)}  \,.
\end{align}
To simplify notation, we will henceforth use $\Gamma$ instead of $\Gamma_{\rm CFI}$.

The branch cut for $\sqrt{\Gamma^2+\alpha^2}$ is chosen to run along the imaginary axis from $-i \Gamma$ to $i\Gamma$. In practice, this means that this function is evaluated as
\begin{equation}
    \sqrt{\Gamma^2+\alpha^2} = \vert \alpha - i\Gamma \vert^{1/2} \vert \alpha + i\Gamma \vert^{1/2} e^{i(\psi^+ +\psi^-)/2} \,.
\end{equation}
where $\psi^\pm = \mathrm{arg}\left(\alpha \mp i\Gamma \right) \in [-\pi/2, 3\pi/2[$. 
The Bromwich contour $\mathcal{C}_\alpha$ passes below the lower branch point at $\alpha = -i\Gamma$. When $\tau <0$, we can close the integration contour to the lower half-plane; since no singularity is crossed, one has $\delta n_p(\tau <0,\xi) = 0$ by virtue of Cauchy's theorem.

We apply the method of steepest descent to obtain the asymptotic behavior of $\delta n_b(\tau,\xi)$ for $\Gamma_{\rm CFI }\tau > 1$. The saddle points are defined by $h_\pm' (\alpha;\theta) = 0$, that is,
\begin{equation}
    \alpha^2 \sqrt{\Gamma^2+\alpha^2} = \pm \Gamma^2 \theta \,.
\end{equation}
Squaring both sides of this equation and defining $z=(\alpha/\Gamma)^2$ and $C=\theta/\Gamma$ gives
\begin{equation}
    z^3+z^2-C^2 = 0 \,.
    \label{eq:P_z}
\end{equation}
The roots of this polynomial are~\cite{Abramowitz_1972}
\begin{align}
    z_1 &= -\frac{1}{3} - \frac{1-i\sqrt{3}}{6} A^{1/3} - \frac{1+i\sqrt{3}}{6} A^{-1/3} \,, \label{eq:z1}\\
    z_2 &= -\frac{1}{3} - \frac{1+i\sqrt{3}}{6} A^{1/3} - \frac{1-i\sqrt{3}}{6} A^{-1/3} \,, \label{eq:z2}\\
    z_3 &= \frac{1}{3}\left(-1 + A^{1/3} + A^{-1/3} \right)\,, \label{eq:z3}
\end{align}
where
\begin{equation}
    A = \frac{1}{2}\left(-2 + 27C^2 + 27C^2 \sqrt{C^2-4/27} \right) \,.
    \label{eq:A}
\end{equation}
In the above, $A^{1/3}$ and $A^{-1/3}$ stand, respectively, for the principal branches of the cubic and inverse cubic roots.

An important observation is that the solutions $\{z_l\}_{l=1,2,3}$ are all real when $C \equiv \xi/\Gamma \tau$ is lower than the critical value $C_\mathrm{cr} = 2/\sqrt{27} \simeq 0.3849$ \cite{Abramowitz_1972}. Further inspection then shows that $z_1,z_2 <0$ while $z_3 > 0$, and hence the saddle points of $h_-$ are
\begin{align}
    &\alpha_{s0}^- =- \Gamma \sqrt{z_3} \,, \\ 
    &(\alpha_{s1}^-, \alpha_{s3}^-)=(-i\Gamma \sqrt{z_2}, i\Gamma \sqrt{z_2}) \,, \\
    &(\alpha_{s2}^-, \alpha_{s4}^-)=(-i\Gamma \sqrt{z_1}, i\Gamma \sqrt{z_1}) \,.
\end{align}
Note that the saddle points at $\{\alpha_{sl}^-\}_{l=1,4}$ are located on the principal (right-hand) sheet of $\sqrt{\Gamma^2 + \alpha^2}$, which assumes $\mathrm{arg} (\pm i \sqrt{z_{1,2}}- i\Gamma) = -\pi/2$. The saddle points of $h_+$ are symmetric to those of $h_-$ with respect to the imaginary axis: $\alpha_{s0}^+ =-\alpha_{s0}^-$ and $\{\alpha_{sl}^+\}_{l=1,4}= \{\alpha_{sl}^-\}_{l=1,4}$ with $\mathrm{arg} (\alpha_{sl=1,4}^+- i \Gamma) = 3\pi/2$. This means that $\{\alpha_{sl}^+\}_{l=1,4}$ lie on the left-hand sheet of $\sqrt{\Gamma^2 + \alpha^2}$. 

In the case where $C > C_\mathrm{cr}$, one has $z_1=z_2^\star \in \mathbb{C}$ and $z_3 \in \mathbb{R}^+$. Then $h_-$ and $h_+$ each admits three saddle points: one is real ($\alpha_{s0}^\pm = \pm \Gamma \sqrt{z_3} \in \mathbb{R}^\pm$) while the two others are complex conjugates ($\alpha_{s5}^- = \Gamma \sqrt{z_2}$, $\alpha_{s6}^- = \Gamma \sqrt{z_1} = \alpha_{s5}^{-\star}$, $\alpha_{s5}^+ = -\Gamma \sqrt{z_1} = - \alpha_{s6}^-$, $\alpha_{s6}^+ = -\Gamma \sqrt{z_2} = \alpha_{s5}^{+\star} = - \alpha_{s5}^-$). 

From those properties, we anticipate, as demonstrated below, that $\delta n_p$ will exhibit distinct spatiotemporal behaviors in the $C \ll C_\mathrm{cr}$ and $C \gg C_\mathrm{cr}$ limits.

\subsection{Case of $C \ll C_\mathrm{cr}$}

\begin{figure}
    \centering
    \includegraphics[width=0.45\textwidth]{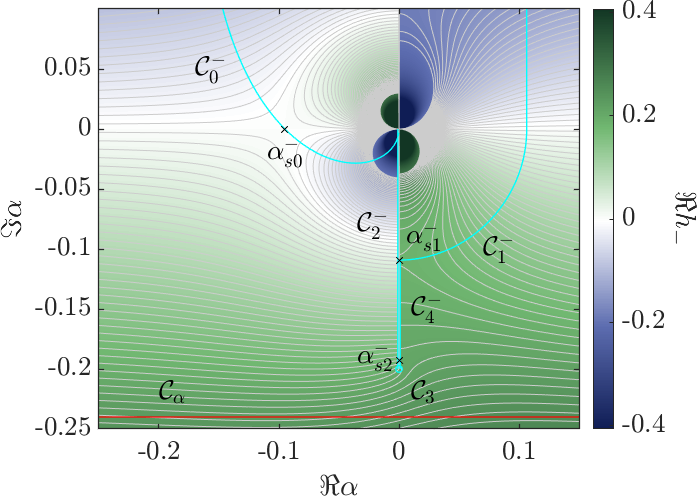}
    \includegraphics[width=0.45\textwidth]{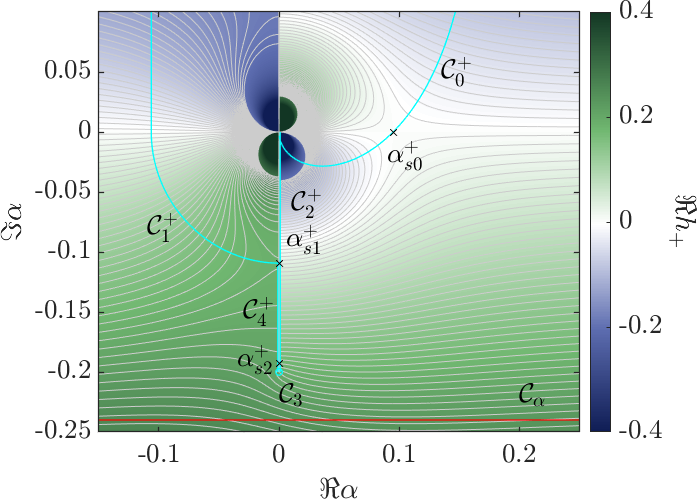}
    \caption{Maps of (top) $\Re h_-(\alpha;\theta)$ and (bottom) $\Re h_+(\alpha;\theta)$ in the complex $\alpha$-plane for $\Gamma=0.2$ and $\theta= 0.05$ ($C = 0.25 < C_\mathrm{cr}$). Grey curves are isocontours of $\Re h_\pm(\alpha;\theta)$. The original Bromwich contour is plotted in red while the combinations of contours used for the asymptotic expansions of $I_\pm$ are plotted in cyan.}
    \label{fig:integ_contours_small_C}
\end{figure}

To make analytical progress, we first consider the $C \ll 1$ limit, in which case the saddle points of $h_\pm$ can be approximated as
\begin{align}
    &\alpha_{s0}^\pm \simeq \pm \sqrt{\Gamma \theta} (1-\theta/4\Gamma) \,,\\
    &\alpha_{s1}^\pm = - \alpha_{s3}^\pm \simeq -i\sqrt{\Gamma \theta} (1+\theta/4\Gamma) \,,\\
    &\alpha_{s2}^\pm = -\alpha_{s4}^\pm \simeq - i\Gamma (1-\theta^2/2\Gamma^2) \,.
\end{align}
To obtain the leading term of the time-asymptotic expansion of $I_-$ in Eq.~\eqref{eq:np_inverse_LT}, we replace the original Bromwich contour, $\mathcal{C}_\alpha$, by the combination of contours $\mathcal{C}_0^- \cup \mathcal{C}_2^- \cup \mathcal{C}_3 \cup \mathcal{C}_4^- \cup \mathcal{C}_1^-$ as depicted in Fig.~\ref{fig:integ_contours_small_C}(a). In detail, $\mathcal{C}_0^-$, which goes from $-\infty + i\infty$ to $\alpha = 0$, coincides with the path of steepest descent near the saddle point $\alpha_{s0}^-$ and is such that $\Re h_-(\alpha;\theta) < \Re h_-(\alpha_{s0}^-;\theta)$ everywhere along it (i.e., it runs along a ``valley'' of $\alpha_{s0}^-$). The contour $C_2^-$ goes down the branch cut, on its left side from $\alpha = 0$ to the branch point $\alpha = -i\Gamma$. The contour $\mathcal{C}_3$ loops around the branch point, while $C_4^-$ runs to the right of the branch cut, from $-i\Gamma$ to the saddle point $\alpha_{s1}^-$. Finally, $\mathcal{C}_1^-$ goes down the valley from $\alpha_{s1}^-$ to $\infty + i\infty$, following the path of steepest descent in the vicinity of $\alpha_{s1}^-$. $\mathcal{C}_0^-$ and $\mathcal{C}_1^-$ are Olver-type contours, and therefore asymptotically equivalent to the steepest descent contours from their respective saddle points \cite{Oughstun_Book_2009}.

The integral along $\mathcal{C}_0^-$ can be evaluated via the method of steepest descent \cite{Oughstun_Book_2009}:
\begin{equation}
    I_0^- 
    \simeq \frac{1}{4\pi} g(\alpha_{s0}^-) \left \vert \frac{2\pi }{ \tau h_-''(\alpha_{s0}^{-};\theta)}\right \vert^{1/2} e^{\tau h_-(\alpha_{s0}^-;\theta) +i\psi_{\rm sd 0}^- + i\pi/2} \,.
    \label{eq:sdp0m} 
\end{equation}
At the saddle point, we have approximately
\begin{align}
    &g(\alpha_{s0}^-) \simeq \frac{1}{
    \Gamma^{3/2} \theta^{1/2}} \,, \label{eq:g_0m} \\
    &h_-(\alpha_{s0}^-;\theta) \simeq -2i \sqrt{\Gamma \theta} \,, \label{eq:h_0m} \\
    & h''_-(\alpha_{s0}^-;\theta) \simeq \frac{-2i}{\sqrt{\Gamma \theta}} \label{eq:d2h_0m} \,.
\end{align}
To obtain Eq.~\eqref{eq:d2h_0m}, we have made use of $h''_\pm (\alpha;\theta) = \pm i\frac{\theta \Gamma^2}{\alpha^3 (\Gamma^2 + \alpha^2)^{3/2}}(2\Gamma^2 + 3\alpha^2)$ together with $\frac{1}{\sqrt{\Gamma^2 + \alpha_s^\pm}} = \pm \frac{\alpha_s^{\pm 2}}{\Gamma^2 \theta}$ ($\alpha_s^\pm$ denotes any saddle point of $h_\pm$), so that $h''_\pm (\alpha_s^\pm;\theta) = i\frac{ \alpha_s^{\pm 3}}{(\Gamma^2 \theta)^2}(2\Gamma^2 + 3\alpha_s^{\pm 2})$.
As a result, the steepest descent path from $\alpha_{s0}^-$ makes an angle of 
\begin{equation}
    \psi_{\rm sd0}^- = -\frac{\pi}{2} - \frac{1}{2}\mathrm{arg}[h''_-(\alpha_{ s0}^-;\theta)] = -\pi/4
    \label{eq:psi_sd0m}
\end{equation}
with the real axis. Substituting Eqs.~\eqref{eq:g_0m}-\eqref{eq:psi_sd0m} into Eq.~\eqref{eq:sdp0m} and further simplifying,
\begin{equation}
    I_0^- \simeq \frac{1}{4\sqrt{\pi}} \frac{e^{-2i \sqrt{\Gamma \tau \xi} + i\pi/4}}{\Gamma (\Gamma \tau \xi)^{1/4}} \,,
    \label{eq:I0_m}
\end{equation}
which corresponds to a nongrowing oscillatory term, hence of negligible contribution to the long-timescale evolution of $\delta n_p(\tau,\xi)$. 

The line integral over $\mathcal{C}_2^-$ can be converted into a real integral by using the expressions
\begin{align}
    &\alpha = -i(\Gamma-\rho) \,, 
    \alpha - i \Gamma = (2\Gamma-\rho)e^{3i\pi/2} \,, \nonumber\\
    &\alpha +i \Gamma = \rho e^{i\pi/2} \,, 
    \sqrt{\Gamma^2 + \alpha^2} = -\sqrt{\rho(2\Gamma- \rho)} \,
    \label{eq:I2_m_change_var}
\end{align}
with $0 < \rho < \Gamma$.
Substitution of these results into the integral over $\mathcal{C}_2^-$ gives
\begin{align}
    I_2^- 
    &=-\frac{i e^{\Gamma \tau}}{4\pi}  \int_0^\Gamma \frac{d\rho}{\sqrt{\rho(2\Gamma-\rho)}} \frac{e^{-\tau [\rho + \frac{\theta}{\Gamma-\rho}\sqrt{\rho(2\Gamma-\rho)} ]}}{(\Gamma-\rho)} \,.
    \label{eq:I2_m}
\end{align}
Although it would be possible to approximate this integral by expanding its integrand around $\rho = 0$, there is no need to do so in order to evaluate $\delta n_p(\tau,\xi)$ since, as will be shown below, $I_2^-$ turns out to be canceled by a similar line integral involving $h_+$.

Turning to the loop integral ($I_3^-$) around $\alpha = -i \Gamma$, it is easy to show that it vanishes by expressing $\alpha = -i\Gamma + \rho e^{i\psi}$ (with $-3\pi/2 < \psi < \pi/2$) and making $\rho$ tend to zero.

As for $I_2^-$, the line integral over $\mathcal{C}_4^-$ can be recast as a real integral by means of the relationships
\begin{align}
    &\alpha = -i(\Gamma - \rho)\,,
    \alpha- i \Gamma = (2\Gamma - \rho)e^{-i\pi/2} \,,\nonumber \\
    &\alpha- i \Gamma = \rho e^{i\pi/2} \,, 
    \sqrt{\Gamma^2 + \alpha^2} = \sqrt{\rho (2\Gamma - \rho)} \,,
    \label{eq:I4_m_change_var}
\end{align}
so that
\begin{align}
    I_4^- 
    &=-\frac{i e^{\Gamma \tau}}{4\pi}  \int_0^{\Gamma + \Im \alpha_{s1}^-} \frac{d\rho}{\sqrt{\rho(2\Gamma-\rho)}} \frac{e^{-\tau [\rho - \frac{\theta}{\Gamma-\rho}\sqrt{\rho(2\Gamma-\rho)} ]}}{(\Gamma-\rho)} \,.
    \label{eq:I4_m}
\end{align}
Again, it is unnecessary to further transform this integral as it will be canceled by a similar one involving $h_+$.

As for the integral over $\mathcal{C}_0^-$, the asymptotic behavior of the integral over $C_1^-$ can be obtained by the saddle-point expansion given in Eq.~\eqref{eq:sdp0m}. To this end, we use the approximations
\begin{align}
    &g(\alpha_{s1}^-) \simeq -\frac{i}{\Gamma^{3/2} \theta^{1/2}} \left(1+\frac{\theta}{4\Gamma} \right) \,,\\
    &h_-(\alpha_{s1}^-;\theta) \simeq 2\sqrt{\Gamma \theta} \left(1-\frac{\theta}{4\Gamma} \right)
    \label{eq:sdp1m}
\end{align}
Moreover, 
\begin{equation}
    h''_-(\alpha_{s1}^-;\theta) \simeq -\frac{2}{\sqrt{\Gamma \theta}}\left(1-\frac{3\theta}{4\Gamma} \right) \,,
\end{equation}
and hence the direction of steepest descent from $\alpha_{s1}^-$ is
\begin{equation}
    \psi_{\rm sd1}^- = \pi/2 - \frac{1}{2} \mathrm{arg}[h''_-(\alpha_{s1}^-;\theta)] = 0 \,.
\end{equation}
Combining these results, and noting that the integral runs only over half the steepest descent path, one finds
\begin{align}
    I_1^- 
    &\simeq -\frac{(1+3\theta/8\Gamma) }{8\sqrt{\pi}} \frac{e^{2 \sqrt{\Gamma \tau \xi}(1-\theta/4\Gamma)}}{\Gamma (\Gamma \tau \xi)^{1/4}} \,.
    \label{eq:I1_m}
\end{align}

We resort to the same procedure to evaluate $I_+$. Specifically, we break the Bromwich contour into the contours $\mathcal{C}_1^+ \cup \mathcal{C}_4^+ \cup \mathcal{C}_3 \cup \mathcal{C}_2^+ \cup \mathcal{C}_0^+$ as indicated in Fig.~\ref{fig:integ_contours_small_C}(b).  These contours being symmetric to their counterparts $\left\{\mathcal{C}_l^-\right\}$ with respect to the imaginary axis, it follows that
\begin{align}
    &I_0^+ \simeq \frac{1}{4\sqrt{\pi}} \frac{e^{-2i \sqrt{\Gamma \tau \xi} + 3i\pi/4}}{\Gamma (\Gamma \tau \xi)^{1/4}} \,, \\
    &I_1^+ \simeq \frac{(1+3\theta/8\Gamma) }{8\sqrt{\pi}} \frac{e^{2 \sqrt{\Gamma \tau \xi}(1-\theta/4\Gamma)}}{\Gamma (\Gamma \tau \xi)^{1/4}} \,, \\
    &I_2^+ = 
     I_2^- \,, \\
    &I_3^+  = 0 \,, \\
    &I_4^+ = 
    I_4^- \,.
\end{align}

Finally, since $\delta n_p(\tau, \xi) = \sum_{l=0}^4 (I_l^+ - I_l^-)$, the $I_2^\pm$ and $I_4^\pm$ integrals cancel out. Because the $I_0^\pm$ integrals are nongrowing and $I_3^\pm = 0$, the dominant term is given by $I_1^+-I_1^-= 2I_1^+$. Hence, one obtains to leading order in $\Gamma_{\rm CFI}\tau > 1$
\begin{align}
    \delta n_p (\tau, \xi) \simeq \frac{(1+3\xi/8\Gamma \tau) }{4\sqrt{\pi}} \frac{e^{2 \sqrt{\Gamma \tau \xi}(1-\xi/4\Gamma \tau)}}{\Gamma (\Gamma \tau \xi)^{1/4}} \,.
    \label{eq:np_small_C}
\end{align}
For $0 < \xi/\Gamma \tau \ll 1$, this expression recovers the asymptotic solution \eqref{eq:sol_strong_qsa_short_asymp} to Eq.~\eqref{eq:CFI_st_master_front}, valid in the limit $\partial_\xi^2 \gg 1$ [see Sec.~\ref{sec:transverse_modes}]. This result could have been intuited noting that in this limit, the dominant integrals $I_1^\pm$ can be approximated by expanding their respective integrand assuming $\alpha^2 \ll \Gamma^2 $, that is, by taking $\sqrt{\Gamma^2 + \alpha^2} \sim \Gamma$ in $I_1^-$ and $\sqrt{\Gamma^2 + \alpha^2} \sim -\Gamma$ in $I_1^+$. As a consequence,  
\begin{align}
    \delta n_p(\tau,\xi) &\simeq -\frac{i}{4\pi \Gamma} \int_{\mathcal{C}_1^+ \cup \mathcal{C}_1^-} \frac{d\alpha}{\alpha}\, e^{\tau (\alpha - \Gamma \theta/\alpha)} \nonumber \\
    &\simeq \frac{1}{2\Gamma} I_0 \left(2 \sqrt{\Gamma \tau \xi} \right) \,.
\end{align}

\subsection{Case of $C \gg C_\mathrm{cr}$}

\begin{figure}[t]
    \centering
    \includegraphics[width=0.45\textwidth]{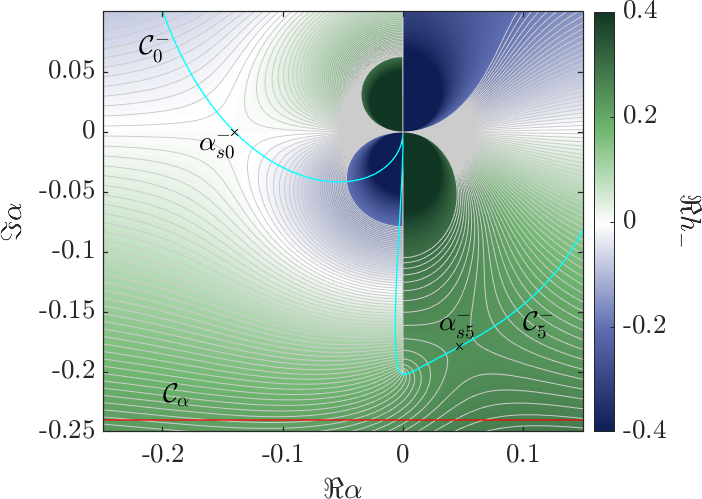}
    \includegraphics[width=0.45\textwidth]{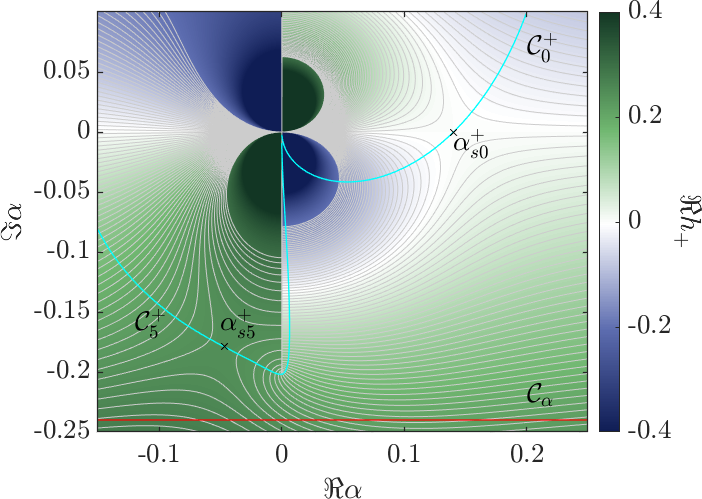}
    \caption{Maps of (a) $\Re h_-(\alpha;\theta)$ and (b) $\Re h_+(\alpha;\theta)$ in the complex $\alpha$-plane for $\Gamma=0.2$ and $\theta=0.12$ ($C=0.6 > C_\mathrm{cr}$). Grey curves are isocontours of $\Re h_\pm(\alpha;\theta)$. The original Bromwich contour is plotted in red while the combinations of contours used for the asymptotic expansions of $I_\pm$ are plotted in cyan.}
    \label{fig:integ_contours_large_C}
\end{figure}

In this parameter range, the non-real saddle points of $h_\pm$, which will be shown to dominate the system dynamics, move off the imaginary axis as illustrated in Fig.~\ref{fig:integ_contours_large_C} for $\theta = 0.6$. For analytical tractability, we assume that $C \gtrsim 1$, in which case the roots of Eq.~\eqref{eq:P_z} can be expanded as
\begin{align}
    &z_1 \simeq - e^{-i\pi/3} C^{2/3} - \frac{1}{3} - \frac{e^{i\pi/3}}{9C^{2/3}} \,, \\
    &z_2 \simeq - e^{i\pi/3} C^{2/3} - \frac{1}{3} - \frac{e^{-i\pi/3}}{9C^{2/3}} \,, \\
    &z_3 \simeq  C^{2/3} - \frac{1}{3} + \frac{1}{9C^{2/3}} \,.
\end{align}
There follow the approximate expressions for the saddle points of $h^\pm$:
\begin{align}
    &\alpha_{s0}^+ = -\alpha_{s0}^- \simeq \Gamma \left(C^{1/3} -\frac{1}{6 C^{1/3}} +\frac{1}{24 C} \right) \,, \\
    &\alpha_{s5}^+ = \alpha_{s6}^{+ \star} \simeq \Gamma \left(-e^{i\pi/3} C^{1/3} + \frac{e^{-i\pi/3}}{6C^{1/3}} + \frac{1}{24 C} \right) \,,\\
    &\alpha_{s5}^- = \alpha_{s6}^{- \star} \simeq \Gamma \left(e^{-i\pi/3} C^{1/3} - \frac{e^{i\pi/3}}{6C^{1/3}} - \frac{1}{24 C} \right) \,.
\end{align}

As shown in Figs.~\ref{fig:integ_contours_large_C}(a,b), the original integration contour $\mathcal{C}_\alpha$ for $h^\pm$ can be deformed to the contours $\mathcal{C}_0^\pm \cup \mathcal{C}_5^\pm$, where $\mathcal{C}_0^\pm$ and $\mathcal{C}_5^\pm$ follow the steepest descent paths from the saddle points $\alpha_{s0}^\pm$ and $\alpha_{s5}^\pm$ given above. Since $\Re h^\pm (\alpha_{s5}^\pm;\theta) > \Re h^\pm (\alpha_{s0}^\pm;\theta) = 0$, the dominant contributions come from $\alpha_{s5}^\pm$. To evaluate them, the following approximations are needed:
\begin{align}
    g(\alpha_{s5}^+) = g(\alpha_{s5}^-)^\star \simeq - \frac{e^{i\pi/3}}{\Gamma^2 C^{2/3}} \left( 1 + \frac{e^{i\pi/3}}{6C^{2/3}} \right)\,,
\end{align}
\begin{align}
    h_+(\alpha_{s5}^+;\theta) &= h_-(\alpha_{s5}^-;\theta)^\star \nonumber \\
    &\simeq i\Gamma \left(C - \frac{3}{2}e^{i\pi/3} C^{1/3} + \frac{e^{-i\pi/3}}{8C^{1/3}}\right) \,, 
\end{align}
\begin{align}
    &h''_+(\alpha_{s5}^+;\theta) = h''_-(\alpha_{s5}^-;\theta)^\star \simeq \frac{3 e^{-5i\pi/6}}{\Gamma C^{1/3}} \left(1+\frac{e^{i\pi/3}}{6C^{2/3}} \right)\,.
\end{align}
From the latter relation we deduce
\begin{align}
    &\vert h''_\pm (\alpha_{s5}^\pm;\theta) \vert \simeq \frac{3}{\Gamma C^{1/3}}\left(1+\frac{1}{12C^{2/3}} \right) \,,\\ 
    &\mathrm{arg} \left[h''_\pm (\alpha_{s5}^\pm;\theta)\right] = \mp \frac{5\pi}{6} \pm \arctan \left( \frac{\sqrt{3}}{12C^{2/3}+1} \right) \,,
\end{align}
so that the slope angles of the steepest descent paths at $\alpha_{s5}^\pm$ are
\begin{equation}
    \phi_{\rm sd5}^\pm \simeq \mp \pi/2 - \frac{1}{2} \mathrm{arg} \left[h''_\pm (\alpha_{s5}^\pm ;\theta)\right] \simeq \mp \frac{\pi}{12} \mp \frac{1}{8\sqrt{3} C^{2/3}} \,.
\end{equation}
Now, plugging the above expressions into Eq.~\eqref{eq:sdp0m} yields
\begin{widetext}
\begin{align}
   &I_5^\pm = \frac{i}{4\pi}\int_{\mathcal{C}_5^\pm} d\alpha\, g(\alpha) e^{\tau h_\pm(\alpha; \theta)} \,\nonumber 
   \simeq \pm \frac{1}{4 \Gamma} \left(\frac{2}{3\pi \xi}\right)^{1/2} \left[1 + \frac{1}{6 C^{2/3}}\left(\frac{1}{4} \pm i\frac{\sqrt{3}}{2} \right) \right] e^{\pm i \xi + \frac{3}{4}(\sqrt{3} \mp i)(\Gamma \tau)^{2/3} \xi^{1/3} \mp i\pi/4} \,.
\end{align}
To make a connection with the ultrarelativistic OTSI~\cite{San_Miguel_Claveria_PRR_2022}, we introduce its temporal growth rate $\Gamma_{\rm OTSI} = \frac{\sqrt{3}}{2^{4/3}} \Gamma^{2/3}$. Therefore, when $C \gg 1$, the time-asymptotic expansion of $\delta n_p(\tau,\xi)$ can be written as
\begin{align}
    \delta n_p(\tau,\xi) &\simeq I_5^+ - I_5^- \simeq \frac{H(\tau) H(\xi)}{4 \Gamma} \left(\frac{2}{3\pi \xi}\right)^{1/2} \left\{\left[1 + \frac{1}{6 C^{2/3}}\left(\frac{1}{4} + i\frac{\sqrt{3}}{2} \right) \right] e^{ i \xi + \frac{\sqrt{3}}{2^{2/3}}(\sqrt{3}-i)\Gamma_{\rm OTSI} \tau^{2/3} \xi^{1/3} - i\pi/4} 
    + c.c. \right\} \,.
    \label{eq:np_large_C}
\end{align}
\end{widetext}
We thus recover the asymptotic spatiotemporal behavior of OTSI as obtained in the slowly-varying-envelope approximation \cite{San_Miguel_Claveria_PRR_2022}.

\subsection{{Case of $C = C_\mathrm{cr}$}}

In the vicinity of the critical point $C_\mathrm{cr}$, we have
\begin{align}
    &A^{1/3} \simeq 1 + \frac{2}{3^{1/4}} \sqrt{C-C_\mathrm{cr}} + \frac{2}{\sqrt{3}} \left(C-C_\mathrm{cr} \right) \,, \\
    &A^{-1/3} \simeq 1 - \frac{2}{3^{1/4}} \sqrt{C-C_\mathrm{cr}} + \frac{2}{\sqrt{3}} \left(C-C_\mathrm{cr} \right) \,.
\end{align}
As a consequence, the relevant saddle points can be approximated as
\begin{align}
    &\alpha_{\rm s1}^\pm \simeq -i \sqrt{\frac{2}{3}} \Gamma \left(1-\frac{3^{1/4}}{2} \sqrt{C_\mathrm{cr}-C} \right) \,, \\
    &\alpha_{\rm s2}^\pm \simeq -i \sqrt{\frac{2}{3}} \Gamma \left(1+\frac{3^{1/4}}{2} \sqrt{C_\mathrm{cr}-C} \right) \,
\end{align}
for $C<C_\mathrm{cr}$, and
\begin{equation}
    \alpha_{\rm s5}^\pm \simeq -\sqrt{\frac{2}{3}} \Gamma \left( \pm \frac{3^{1/4}}{2}\sqrt{C-C_\mathrm{cr}} + i \right)
\end{equation}
for $C>C_\mathrm{cr}$.

It is possible to obtain an asymptotic expansion of $I_1^\pm$ that is a continuous function of $C$ about $C_\mathrm{cr}$ \cite{Felsen_Marcuwitz_Book_2009, Oughstun_Book_2009}. Here, for simplicity, we restrict our analysis to the behavior of the Green's function at the critical point $C=C_\mathrm{cr}$ or, equivalently, at $\theta = \theta_\mathrm{cr} \equiv C_\mathrm{cr} \Gamma$. The two saddle points $\alpha_{\rm s1}^\pm (\theta)$ and $\alpha_{\rm s2}^\pm (\theta)$ then coalesce into a single saddle point of second order, $\alpha_{\rm s1}^\pm (\theta_\mathrm{cr}) \simeq -i \sqrt{2/3} \Gamma$, such that $h_\pm'(\alpha_{\rm s1}^\pm ; \theta_\mathrm{cr}) = h_\pm''(\alpha_{\rm s1}^\pm ; \theta_\mathrm{cr}) = 0$. An asymptotic approximation of the integral $I_1^-$ is then \cite{Felsen_Marcuwitz_Book_2009}:
\begin{equation}
    I_1^- \simeq \frac{i \Gamma(1/3)}{12\pi} g(\alpha_{\rm s1}^-) \left \vert \frac{6}{\tau h_-'''(\alpha_{\rm s1}^-; \theta_\mathrm{cr})} \right \vert^{\frac{1}{3}} e^{\tau h_-(\alpha_{\rm s1}^-; \theta_\mathrm{cr}) + i\psi_{\rm sd1}^-} \,,
    \label{eq:I1m_o2}
\end{equation}
where $\Gamma(1/3) \simeq 2.6789$ denotes the gamma function and $\psi_{\rm sd1}^-$ is the slope angle of the steepest descent path from $\alpha_{\rm s1}^-$, given by
\begin{equation}
    \psi_{\rm sd1}^- = \frac{\pi}{3} - \frac{1}{3}\mathrm{arg}[h_-'''(\alpha_{\rm s1}^-;\theta_\mathrm{cr})] \,.
\label{eq:psi_sd_o2}
\end{equation}
In addition, combining
\begin{equation}
    h_-'''(\alpha;\theta) = 2i \Gamma^2 \theta \frac{( 4 \alpha^4 + 5\Gamma^2 \alpha^2 + 2\Gamma^4)}{\alpha^4(\Gamma^2 + \alpha^2)^{5/2}}
\end{equation}
and $\sqrt{\Gamma^2 + (\alpha_{\rm s1}^{-})^2} = - \Gamma^2 \theta/(\alpha_{\rm s1}^{-})^2$, one finds
\begin{equation}
    h_-'''(\alpha_{\rm s}^-;\theta) = - \frac{3i (\alpha_{\rm s}^-)^6}{(\Gamma^2 \theta)^4} \left[4 (\alpha_{\rm s}^-)^4 + 5 \Gamma^2 (\alpha_{\rm s}^-)^2 + 2 \Gamma^4 \right] \,,
\end{equation}
and hence, after substituting $\alpha_{\rm s1}^{-} \simeq -i \sqrt{2/3}\Gamma$,
\begin{equation}
    h_-'''(\alpha_{\rm s1}^-;\theta_\mathrm{cr}) \simeq \frac{32 i}{81} \frac{\Gamma^2}{\theta_{\rm cr}^4} \,, 
\end{equation}
so that $\mathrm{arg}[h_-'''(\alpha_{\rm s1}^-;\theta_\mathrm{cr})] = \pi/2$. Equation~\eqref{eq:psi_sd_o2} then yields
$\psi_{\rm sd1}^- = \pi/6$. We also have $h(\alpha_{\rm s1}^{-};\theta_{\rm cr}) \simeq \frac{4}{3}\sqrt{\frac{2}{3}} \Gamma$. 

We now insert the previous expressions into Eq.~\eqref{eq:I1m_o2} to obtain  
\begin{equation}
    I_1^- \simeq - \frac{6^{7/6} \Gamma(1/3)}{48\pi \Gamma} \left(\frac{\xi}{\Gamma \tau}\right)^{1/3} e^{\frac{4}{3} \sqrt{\frac{2}{3}} \Gamma \tau + i\pi/6} \,.
\end{equation}
The same procedure leads to $I_1^+ = -I_1^{-\star}$, and therefore 
\begin{align}
    \delta n_p(\tau,\xi = C_\mathrm{cr}\Gamma \tau) &\simeq -2 \Re I_1^- \nonumber \\
    &\simeq \frac{3^{1/6} \Gamma(1/3) e^{\frac{4}{3} \sqrt{\frac{2}{3}} \Gamma \tau}}{4\sqrt{2} \pi  \Gamma (\Gamma \tau)^{1/3}} \,.
    \label{eq:np_C_cr}
\end{align}
An observer moving away from the beam front according to $\xi = C_\mathrm{cr}\Gamma \tau$ thus sees an essentially exponential temporal growth at a rate of $(4/3)\sqrt{2/3} \Gamma \simeq 1.09 \Gamma$. 

\begin{figure}[t]
    \centering
    \includegraphics[width=0.45\textwidth]{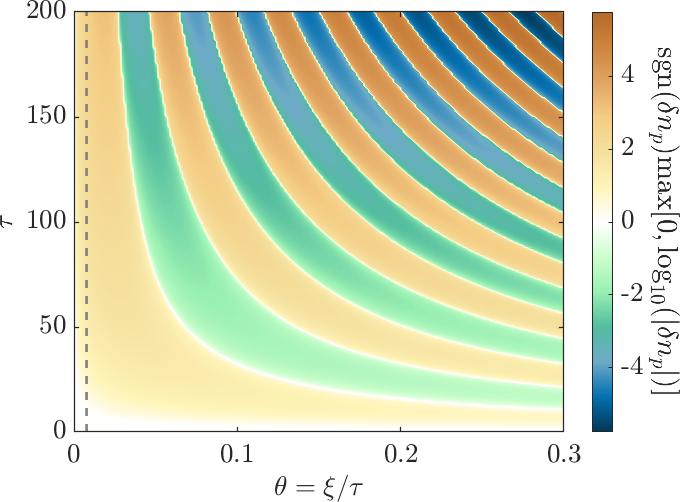} \\[1ex]
    \includegraphics[width=0.45\textwidth]{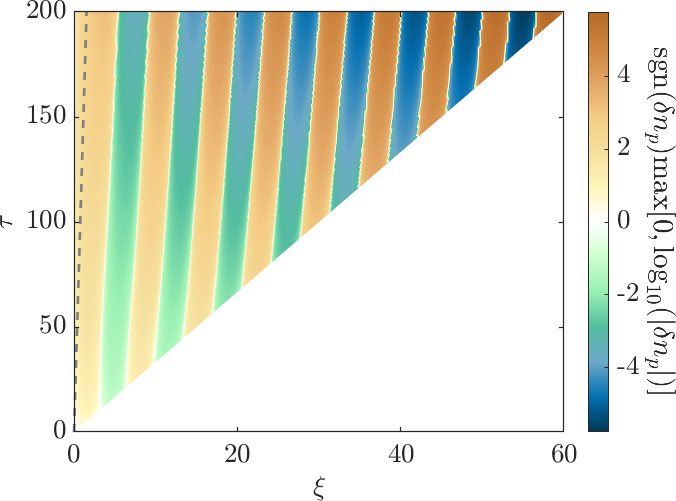}
    \caption{Numerical evaluation of the impulsive solution given by the inverse Laplace transform \eqref{eq:np_inverse_LT} for $\Gamma=0.02$: log-scale representations in the (top) $\theta$--$\tau$ and (bottom) $\xi$--$\tau$ planes. The gray dashed lines indicate the critical point $\theta \equiv \xi/\tau =  \theta_{\rm cr}$.}
    \label{fig:np_exact_Gamma_1e-1}
\end{figure}

\begin{figure}[t]
    \centering
    \includegraphics[width=0.45\textwidth]{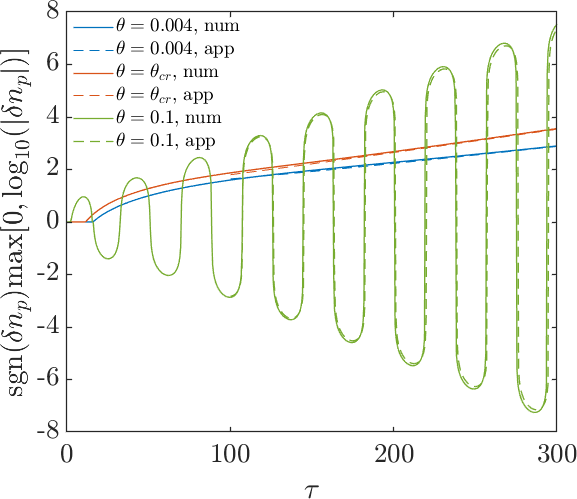}
    \caption{Time evolution of the numerical (``num'', solid curves) and approximate (``app'', dashed curves) impulsive solutions for $\Gamma=0.02$ and different values for $\theta$. For $\theta=0.004$ (blue), $\theta = \theta_{\rm cr} \simeq 0.0077$ (red) and $\theta = 0.1$ (green), the approximations are given by formulas \eqref{eq:np_small_C}, \eqref{eq:np_C_cr} and \eqref{eq:np_large_C}, respectively.}
    \label{fig:np_exact_approx_Gamma_1e-1}
\end{figure}

To conclude, we have demonstrated that the impulsive solution to Eq.~\eqref{eq:CFI_st_master} obtained within the quasistatic approximation captures the asymptotic evolutions of both CFI and OTSI in their respective spatiotemporal domains of predominance. Notably, we have rigorously shown that, for a given transverse wave number, the long-timescale impulsive response of the ultrarelativistic beam-plasma system is governed by
\begin{itemize}
    \item a purely growing, spatiotemporal CFI-type mode, evolving as $\sim e^{2\sqrt{\Gamma \tau \xi}}$ when $\xi/\Gamma \tau \ll C_\mathrm{cr}$;
    \item an oscillatory growing, spatiotemporal OTSI-type mode, evolving as $\sim e^{i \xi + \frac{3}{4}(\sqrt{3}-i)(\Gamma \tau)^{2/3} \xi^{1/3}}$ when $\xi/\Gamma \tau \gg C_\mathrm{cr}$.
\end{itemize}
Interestingly, CFI persists until $\xi/\Gamma \tau = C_{\rm cr}$ where it exhibits purely temporal growth [Eq.~\eqref{eq:np_C_cr}].

This behavior is illustrated in Fig.~\ref{fig:np_exact_Gamma_1e-1}, which presents the exact solutions for $\delta n_p(\theta,\tau)$ (top panel) and $\delta n_p(\xi,\tau)$ (bottom panel) at $\Gamma = 0.02$, obtained via direct numerical integration of Eq.~\eqref{eq:np_inverse_LT}. The dashed line marks the critical point $\theta_\mathrm{cr}$.
The $\theta$ range is restricted to $0 < \theta < 0.3$ because the quasistatic approximation ($\partial_\tau \ll \partial_\xi$) breaks down when $\theta$ approaches unity.

The accuracy of our analytical expansions, Eqs.~\eqref{eq:np_small_C}, \eqref{eq:np_large_C}, and \eqref{eq:np_C_cr}, can be gauged from Fig.~\ref{fig:np_exact_approx_Gamma_1e-1}, which compares the exact and approximate temporal evolutions of $\delta n_p(\tau,\theta)$ for three different values of $\theta$. The expected trends are observed, namely, a nonoscillatory growth for $\theta \le \theta_{\rm cr} \simeq 0.0077$, and an oscillatory growth for $\theta > \theta_{\rm cr}$. The numerical (solid) and analytical (dashed) curves show excellent agreement for $\tau > 100$ ($\Gamma \tau > 2$).

\section{Asymptotic solution of Eq.~\eqref{eq:CFI_st_master} for an extended initial disturbance}
\label{sec:AppendixC}

The impulsive solution, associated with a disturbance localized at $(\xi,\tau)=(0,0)$, is expected to be primarily relevant when the beam crosses a sharp vacuum-plasma boundary. Another situation of interest occurs when the beam is driven directly within the plasma via an external source. In this case, we expect the following initial and boundary conditions to be more appropriate \cite{Pathak_NJP_2015, San_Miguel_Claveria_PRR_2022}:
\begin{align}
    &\delta n_p(\tau=0,\xi) = \delta n_p(\tau,\xi=0) = S \,, \nonumber \\
    &\partial_\tau \delta n_p(\tau=0,\xi) = 0 \,,
    \label{eq:extended_BC}
\end{align}
where $S$ is the initial fluctuation amplitude. 

The double Laplace transform of Eq.~\eqref{eq:CFI_st_master} then gives
\begin{equation}
    \delta n_p(\alpha,\beta) = \frac{S(1-\beta^2)}{\alpha \beta \Big(\beta - \sqrt{\frac{\Gamma^2}{\alpha^2} +1} \Big) \Big(\beta + \sqrt{\frac{\Gamma^2}{\alpha^2} +1} \Big)} \,,
\end{equation}
where we have used $\delta n_p(\alpha,\xi=0) = -iS/\alpha$ and $\delta n_p(\tau=0,\beta=0) = -iS/\beta$. Again, for brevity, we write $\Gamma$ instead of $\Gamma_{\rm CFI}$.

It follows that
\begin{align}
    &\delta n_p(\alpha,\xi) = \frac{S}{2\pi \alpha} \nonumber \\
    &\times \int_{\mathcal{C}_\beta} d\beta\, \frac{(1-\beta^2) e^{i\beta \xi}}{\beta \Big( \beta - \sqrt{\frac{\Gamma^2}{\alpha^2} +1} \Big) \Big(\beta + \sqrt{\frac{\Gamma^2}{\alpha^2} +1} \Big)} \,.
\end{align}
The integrand contains poles at $\beta = 0$ and $\beta = \pm \sqrt{\Gamma^2/\alpha^2+1}$. The residue theorem yields
\begin{align}
    &\delta n_p(\alpha,\xi) = -\frac{iS}{\alpha(\Gamma^2+\alpha^2)} H(\xi) \nonumber \\
    &\times \left[ \alpha^2 + \frac{\Gamma^2 }{2} \left( e^{i\xi \sqrt{\frac{\Gamma^2}{\alpha^2} +1}} + e^{-i\xi \sqrt{\frac{\Gamma^2}{\alpha^2} +1}} \right) \right] \,.
\end{align}
It now remains to evaluate the inverse transform with respect to $\alpha$, which we recast as
\begin{equation}
    \delta n_p(\tau,\xi) = (K + J^{+} + J^{-})\,H(\xi) \,,
    \label{eq:np_ext_general_form}
\end{equation}
where 
\begin{align}
    K &= -\frac{i S}{2\pi} \int_{\mathcal{C}_\alpha} d\alpha \frac{\alpha\, e^{i\alpha \tau}}{\alpha^2 + \Gamma^2}  \,, \label{eq:int_K} \\
    J^\pm &= -\frac{i S \Gamma^2}{4\pi} \int_{C_\alpha} d \alpha \frac{e^{\tau h_\pm(\alpha;\theta)}}{\alpha^2 + \Gamma^2} \,, \label{eq:int_J_pm}
\end{align}
and the functions $h_\pm(\alpha;\theta)$ are defined in Eq.~\eqref{eq:h_pm}.

Applying the residue theorem to the poles at $\alpha = \pm i \Gamma$, the integral $K$ can be exactly solved as
\begin{equation}
    K = S \cosh (\Gamma \tau) H(\tau) \,.
\end{equation}

The integrals $J^\pm$ are similar to the integrals $I^\pm$ considered in Appendix~\ref{sec:AppendixB}, except that the prefactor $g(\alpha)$ is now
\begin{equation}
    g(\alpha) = \frac{1}{\alpha (\alpha^2 + \Gamma^2)} 
\end{equation}
Since $I^\pm$ and $J^\pm$ share the same saddle points, we again distinguish between the regimes of $C \ll C_{\rm cr}$ and $C \gg C_{\rm cr}$, and exploit the analytical results of Appendix~
\ref{sec:AppendixB}.

\subsection{Case of $C \ll C_\mathrm{cr}$}

The Bromwich contour $\mathcal{C}_\alpha$ of $J^{-}$ can be deformed into the combination of contours $\mathcal{C}_0^{-} \cup \mathcal{C}_2^{-} \cup \mathcal{C}_3 \cup \mathcal{C}_4^{-} \cup \mathcal{C}_1^-$, as defined in Fig.~\ref{fig:integ_contours_small_C}. Accordingly, the integral is decomposed as $J^{-} = J_0^{-} + J_2^{-} + J_3^{-} + J_4^{-} + J_1^{-}$. Each of these terms can 
be readily evaluated based on the results of Appendix~\ref{sec:AppendixB}:
\begin{itemize}
    \item $J_0^-$: Following Eqs.~\eqref{eq:sdp0m}--\eqref{eq:I0_m}, the method of steepest descent yields a purely oscillatory term, $J_0^- \propto e^{-2i\sqrt{\Gamma \tau \xi}}$, which is negligible relative to the other integrals.
    \item $J_2^-$: Repeating the calculation of Eqs.~\eqref{eq:I2_m_change_var}--\eqref{eq:I2_m}, this line integral can be simplified to
    \begin{equation}
        J_2^{-} = -\frac{iS\Gamma^2}{4\pi} \int_0^\Gamma du \frac{e^{\tau \left(u-\frac{\theta}{u}\sqrt{\Gamma^2-u^2} \right)}}{u(\Gamma^2 -u ^2)} \,.
    \end{equation}
    \item $J_3^{-}$: To evaluate this loop integral around $\alpha = -i \Gamma$, we make the change of variable $\alpha = - i \Gamma + \rho e^{i\Psi}$ with $-3\pi/2 < \Psi < \pi/2$ and $\rho \to 0$. This gives $h_-(\alpha) \simeq \Gamma$ and $g(\alpha) \simeq -(2\Gamma^2 \rho e^{i\Psi})^{-1}$, so that
    \begin{equation}
        J_3^{-} = -\frac{S}{4}e^{\Gamma \tau} \,. 
    \end{equation}
    \item $J_4^-$: Using Eq.~\eqref{eq:I4_m_change_var}, we obtain
    \begin{equation}
        J_4^{-} = \frac{iS\Gamma^2}{4\pi} \int_{-\Im \alpha_{s1}^{-}}^\Gamma du \frac{e^{\tau \left(u+\frac{\theta}{u}\sqrt{\Gamma^2-u^2} \right)}}{u(\Gamma^2 -u ^2)} \,.
    \end{equation}
    \item $J_1^{-}$: Drawing upon Eqs.~\eqref{eq:sdp1m}--\eqref{eq:I1_m} and using $g(\alpha_{s1}^{-})\simeq i/(\Gamma^{5/2}\sqrt{\theta})$, the method of steepest descent leads to
    \begin{equation}
        J_1^{-} \simeq \frac{S}{8\sqrt{\pi}} \frac{e^{2\sqrt{\Gamma \xi \tau}}}{(\Gamma \xi \tau)^{1/4}} \,.
    \end{equation}
\end{itemize}

Likewise, for the evaluation of $J^{+}$, we deform $\mathcal{C}_\alpha$ as $\mathcal{C}_1^{+} \cup \mathcal{C}_4^{+} \cup \mathcal{C}_3 \cup \mathcal{C}_2^{+} \cup \mathcal{C}_0^+$, and hence split $J^{+} = J_1^{+} + J_4^{+} + J_3^{+} + J_2^{+} + J_0^{+}$. Using the above results, we find
\begin{align}
    J_0^{+} &\propto e^{2i\sqrt{\Gamma \xi \tau}} \,, \\
    J_1^{+} &\simeq \frac{S}{8\sqrt{\pi}} \frac{e^{2\sqrt{\Gamma \xi \tau}}}{(\Gamma \xi \tau)^{1/4}} \simeq J_1^{-} \,, \\
    J_4^{+} &= -\frac{iS \Gamma^2}{4\pi} \int_{-\Im \alpha_{s1}^-}^\Gamma du \frac{e^{\tau \left(u+\frac{\theta}{u}\sqrt{\Gamma^2-u^2} \right)}}{u(\Gamma^2 -u ^2)} = -J_4^{-} \,, \\
    J_3^{+} &= -\frac{S}{4}e^{\Gamma \tau} = J_3^{-} \,, \\
    J_2^{+} &= - \frac{iS\Gamma^2}{4\pi} \int_0^\Gamma du \frac{e^{\tau \left( u-\frac{\theta}{u} \sqrt{\Gamma^2-u^2} \right)}}{u(\Gamma^2 -u ^2)} = -J_2^{-} \,. 
\end{align}

Finally, summing all contributions results in
\begin{align}
    &\delta n_p(\tau,\xi) = K + \sum_{l=0}^4 \left( J_l^- + J_l^+\right) \nonumber \\
    &\simeq S \cosh(\Gamma \tau) + 2\left(J_3^{-} + J_1^{-}\right) \nonumber \\
    &\simeq S \cosh(\Gamma \tau) - \frac{S}{2} e^{\Gamma \tau} + \frac{S}{4\sqrt{\pi}} \frac{e^{2\sqrt{\Gamma \xi \tau}}}{(\Gamma \xi \tau)^{1/4}}
\end{align}
for $(\tau,\xi) > 0$. This yields the $\tau$-asymptotic behavior 
\begin{equation}
    \delta n_p(\tau,\xi) \simeq \frac{S}{4\sqrt{\pi}} \frac{e^{2\sqrt{\Gamma \xi \tau}}}{(\Gamma \xi \tau)^{1/4}} H(\xi) \,.
\end{equation}
Therefore, we recover the dominance of spatiotemporal CFI near the beam front originally predicted by the impulsive solution, owing to the exact cancellation of the $\propto e^{\Gamma \tau}$ terms from $K$ and the $J_3^\pm$ loop integrals. 

\begin{figure}[t]
    \centering
    \includegraphics[width=0.45\textwidth]{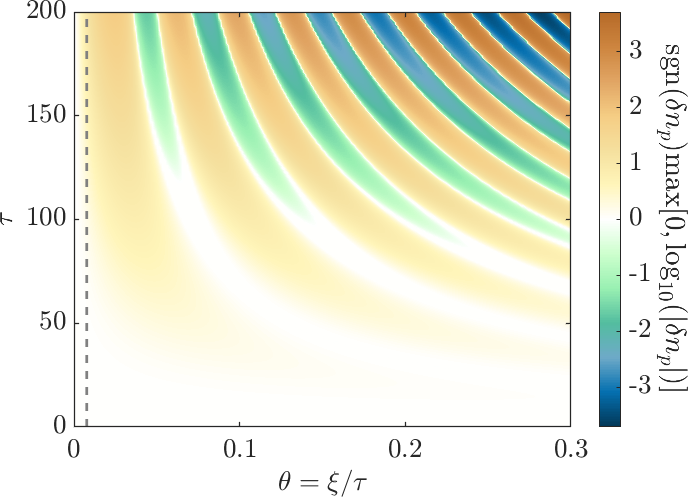} \\[1ex]
    \includegraphics[width=0.45\textwidth]{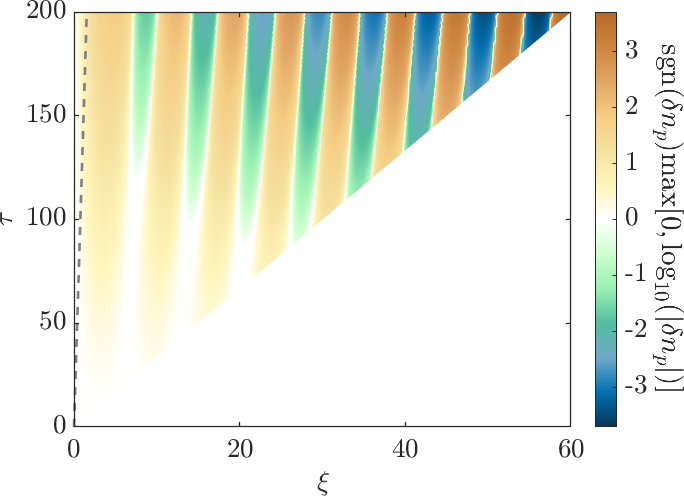}
    \caption{Numerical evaluation of Eq.~\eqref{eq:np_ext_general_form} for an extended initial disturbance with $\Gamma =0.02$: log-scale representations in the (top) $\theta$--$\tau$ and (bottom) $\xi$--$\tau$ planes. The gray dashed lines indicate the critical point $\theta = \theta_{\rm cr}$.}
    \label{fig:np_ext_exact_Gamma_2e-2}
\end{figure}

\subsection{Case of $C \gg C_\mathrm{cr}$}

In Appendix~\ref{sec:AppendixB}, we have demonstrated that the asymptotic behavior of the $I^\pm$ integrals is dominated by the integration along the steepest descent path from the saddle points $\alpha_{s5}^\pm \simeq \mp \Gamma C^{1/3} e^{\pm i \pi/3}$. Naturally, this result also holds for the $J^\pm$ integrals considered here. Given $g(\alpha_{s5}^\pm) \simeq \pm C/(\Gamma \theta^2)$, we obtain
\begin{align}
    J^\pm \simeq J_5^\pm \simeq \frac{S}{2\sqrt{2\pi}} \frac{(\Gamma \tau)^{1/3}}{\xi^{5/6}} e^{\pm i\xi + \frac{3}{4}(\sqrt{3}\mp i)(\Gamma \tau)^{2/3}\xi^{1/3} \mp 7i\pi/12} \,.
\end{align}
Consequently,
\begin{align}
    &\delta n_p(\tau,\xi) \simeq \left( K + J_5^{-} + J_5^{+} \right) H(\xi) \nonumber \\
    &\simeq \bigg\{\frac{S}{2} e^{\Gamma \tau}
    + \frac{S}{2\sqrt{2\pi}} \frac{(\Gamma \tau)^{1/3}}{\xi^{5/6}} \nonumber \\
    &\times \left[ e^{\pm i\xi + \frac{3}{4}(\sqrt{3}\mp i)(\Gamma \tau)^{2/3}\xi^{1/3} \mp 7i\pi/12}  + c.c. \right] \bigg\} H(\xi) \,.
\end{align}
We see that the system's dynamics in this regime is subject to both temporal CFI (first term) and spatiotemporal OTSI (second term). Yet we can expect the latter to prevail over the former when
\begin{equation}
    \frac{3\sqrt{3}}{4}(\Gamma \tau)^{2/3} \xi^{1/3} \gg \Gamma \tau
    \Leftrightarrow C \gg \left(\frac{4}{3\sqrt{3}}\right)^3 \simeq 0.46 \,,
\end{equation}
which is verified by construction in the regime under consideration. Thus, in the presence of an extended initial disturbance of the form \eqref{eq:extended_BC}, the system should behave similarly to the case of a pulse disturbance at the beam front. This is confirmed by the numerical solution of Eq.~\eqref{eq:np_ext_general_form} for $\Gamma = 0.02$ shown in Fig.~\ref{fig:np_ext_exact_Gamma_2e-2}, which exhibits the same qualitative behavior as that observed in Fig.~\ref{fig:np_exact_Gamma_1e-1}.
Note, though, that this result assumes $\xi \lesssim \tau/3$; at larger distances, an extended disturbance will feed temporal OTSI \cite{San_Miguel_Claveria_PRR_2022}, which is not captured by the quasistatic model.

\bibliography{biblio}

\end{document}